\documentclass[10pt,journal,compsoc]{IEEEtran}
\ifCLASSOPTIONcompsoc
  \usepackage[nocompress]{cite}
\else
  \usepackage{cite}
\fi
\ifCLASSINFOpdf
   \usepackage[pdftex]{graphicx}
  \graphicspath{{../pdf/}{../jpeg/}}
  \DeclareGraphicsExtensions{.pdf,.jpeg,.png}
\else
  \usepackage[dvips]{graphicx}
   \graphicspath{{../eps/}}
   \DeclareGraphicsExtensions{.eps}
\fi
\usepackage[cmex10]{amsmath}
\usepackage[ruled,linesnumbered,vlined]{algorithm2e}
\usepackage{array}
\usepackage{mathrsfs}

\usepackage{mdwmath}
\usepackage{mdwtab}

\usepackage{amsfonts,amssymb}

\usepackage{color}

 \usepackage[caption=false,font=footnotesize]{subfig}

\begin{document}
\title{E-Tree Learning: A Novel Decentralized Model Learning Framework for Edge AI}

\author{Lei~Yang,
        Yanyan~Lu,
        Jiannong~Cao,~\IEEEmembership{Fellow,~IEEE,}
        Jiaming~Huang,
        Mingjin~Zhang
\IEEEcompsocitemizethanks{\IEEEcompsocthanksitem L. Yang, Y. Lu and J. Huang are with the School of Software Engineering, South China University of Technology, Guangzhou, China, 510006. \protect\\ E-mail: sely@scut.edu.cn, 201921043977@mail.scut.edu.cn
\IEEEcompsocthanksitem J. Cao and M. Zhang are with the Department of Computing, Hong Kong Polytechnic University, Hong Kong, China. J. Cao is the corresponding author. E-mail: csjcao@comp.polyu.edu.cn, mingjin.zhang@connect.polyu.hk
}
}


\markboth{IEEE Internet of Things Journal, August~2020}%
{Shell \MakeLowercase{\textit{et al.}}: Bare Demo of IEEEtran.cls for Computer Society Journals}

\IEEEtitleabstractindextext{%
\begin{abstract}
Traditionally, AI models are trained on the central cloud with data collected from end devices. This leads to high communication cost, long response time and privacy concerns. Recently Edge empowered AI, namely Edge AI, has been proposed to support AI model learning and deployment at the network edge closer to the data sources. Existing research including federated learning adopts a centralized architecture for model learning where a central server aggregates the model updates from the clients/workers. The centralized architecture has drawbacks such as performance bottleneck, poor scalability and single point of failure. In this paper, we propose a novel decentralized model learning approach, namely E-Tree, which makes use of a well-designed tree structure imposed on the edge devices. The tree structure and the locations and orders of aggregation on the tree are optimally designed to improve the training convergency and model accuracy. In particular, we design an efficient device clustering algorithm, named by KMA, for E-Tree by taking into account the data distribution on the devices as well as the the network distance. Evaluation results show E-Tree significantly outperforms the benchmark approaches such as federated learning and Gossip learning under NonIID data in terms of model accuracy and convergency.

\end{abstract}

}

\maketitle

\IEEEdisplaynontitleabstractindextext

% \fi
\IEEEpeerreviewmaketitle

\IEEEraisesectionheading{\section{Introduction}\label{sec:introduction}}

AI models and algorithms are increasingly used for IoT applications to support intelligent decision making and operation automation. Recently Edge empowered AI, namely Edge AI, has been proposed to support AI model learning and deployment at the network edge closer to the data sources. Transmitting massive data directly from the IoT devices to the cloud for building AI models causes high communication cost, long response time and privacy concerns.

Most existing research on Edge AI, including federated learning \cite{ref_7} and parameter servers \cite{ref_36}, adopt a distributed machine learning framework, where the edge devices separately train their own models using the local data, while a centralized master located on the cloud iteratively coordinates the aggregation and update of the model parameters for the edge devices. The centralized model aggregation faces several drawbacks including performance bottleneck, poor scalability and single point of failure. A decentralized approach is desirable to address the above issues. To our knowledge, only one work, namely Gossip learning \cite{ref_8}, has been reported, which uses a distributed aggregation framework. However, without considering the data distribution on the edge devices in model aggregation, Gossip learning has slow convergence speed and low inference accuracy.

In this article, we propose E-Tree Learning, a novel decentralized model learning framework for Edge AI, which can overcome the shortcomings of the existing works and achieve high performance in terms of convergence speed and inference accuracy. E-Tree makes use of a well-designed hierarchical aggregation structure imposed on the edge devices. More specifically, edge devices are organized into a tree structure, where the leaf nodes represent the learning workers and the non-leaf nodes represent model aggregators. The tree structure and the locations and orders of aggregation on the tree are optimally decided based on the network resources and data distribution on the edge devices. As such, E-Tree can maximize the parallelism in aggregation to speed up the training convergency and improve the inference accuracy.

There are several challenging issues to be addressed. The first issue is how to construct the model aggregation tree. We need to decide which edge devices are grouped together for aggregation and how many layers the tree should have based on the data distribution on the edge devices. The second issue is how to schedule the aggregation operations onto the edge devices. We need to decide where and when each of the model aggregations is performed based on the computation resources on the edge devices, the network topology and communication resources. The third issue is, in solving the first two issues, we need to address the problem that the data is not Independently and Identically Distributed (IID) on edge devices which is a common challenge in model learning.

In particular, we develop an efficient device clustering algorithm, namely KMA, for E-Tree. KMA utilizes K-Means algorithm to group the edge devices according to the data distribution and network distance of the devices. The purpose of KMA is to generate device groups which have as small inter-group difference as possible in the data distribution. KMA outperforms the existing clustering algorithms in networks only taking into account the network distance. The reason is that the data on the edge devices is usually NonIID, clustering only based on the network distance would lead to an un-uniform distribution of data labels among the groups. This leads to a low model accuracy and slow convergency speed, as shown by experiments.

We develop a simulator to compare the performance between E-Tree and two benchmark approaches, i.e., federated learning and Gossip learning. The simulator is developed on top of an open-source benchmark framework \cite{ref_8}. The overall performance results show that E-Tree outperforms the federated learning by 2.4\% in accuracy and Gossip learning by 14.7\% under a NonIID data. Moreover, we evaluate our proposed KMA device clustering algorithm for E-Tree under various network configurations. KMA has obvious better performance than K-Means and Un-uniform KMA, because it controls the inter-group difference in data distribution.

E-Tree learning facilitates distributed learning and parallel aggregation in Edge AI. It can be applied in a wide range of applications, especially in disaster rescue and forest monitoring where access to the cloud is not available. By using the E-Tree learning, the AI models are trained locally and effectively on the edge devices including the IoT devices with certain computation capabilities and the edge gateways deployed within the IoTs. Due to the advantages of E-Tree learning, it would be an important choice for industrial companies to implement the Edge AI solutions. The contribution of this paper are summarized as follows.

\begin{itemize}
 \item{We propose a novel and decentralized model learning framework for Edge AI. To the best of our knowledge, E-Tree is the first configurable decentralized approach that can achieve fast convergency and high model accuracy.}
\item{We design an efficient device clustering algorithm, named by KMA, for E-Tree learning framework. KMA clusters the edge devices at the bottom layer of E-Tree according the data distribution on the devices and network distance, and thus achieves good performance under the NonIID data.}
\item{We develop a simulator and evaluate E-Tree learning framework and our proposed device clustering algorithm. Results show that E-Tree obviously outperforms the two state-of-arts model learning approaches, i.e., federated learning and Gossip learning, specially under the NonIID data.}
\end{itemize}

\section{Model Learning in Edge AI}

In this section, we define the system model of Edge AI, and then describe the objective and challenges for the model learning in Edge AI. Then, we present the motivation of proposing E-Tree learning.

\subsection{System Model and Objectives}

\begin{figure}
  \centering
  \includegraphics[width=3.5in]{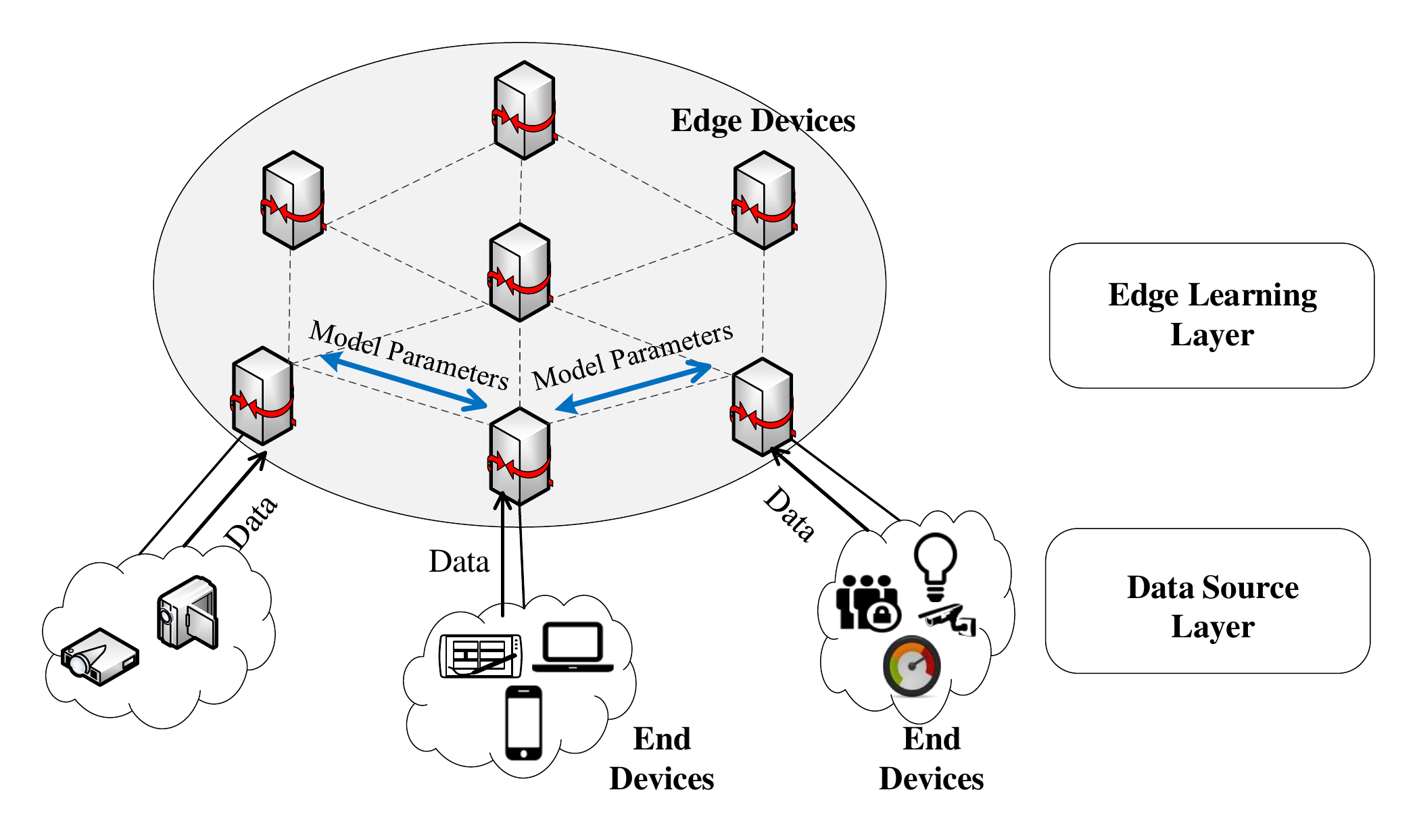}\\
  \vspace{-0.3cm}
  \caption{System model of edge computing}\label{sys_model}
%  \vspace{-0.3cm}
\end{figure}

Fig.\ref{sys_model} shows a generic model of edge computing system. The bottom layer is the data source layer including a large amount of end devices where the data for learning a model are generated. The upper one is the edge computing layer which includes a set of {\itshape edge devices} (or servers) interconnected with each other via particular networks. The {\itshape edge device} is an abstraction of the device/machine that has certain capability. The network connections between the end devices at the data source layer and the edge devices usually have high bandwidth, i.e., directly by using wired cables or short range high data rate communication protocols, while the connections between the edge devices and cloud have much limited bandwidth \cite{add_14}.

The interconnection among the edge devices depends on concrete sceneries. In mobile edge computing, edge devices are MEC servers deployed behind base stations, and they are interconnected by the cellular back-haul networks \cite{add_12}. In industrial IoTs, the edge devices could be smart routers or switches with built-in high performance CPU/GPU cores. The edge devices are interconnected by a wireless mesh network \cite{add_13}. Each edge device collects the data from end devices in its coverage range. Using the data distributed on all the edge devices, the system aims to train a generic model that could be used by every edge device for real-time inference.

Based on the system model above, we consider the following quality attributes/metrics for the model learning. The primary metric is the model accuracy which is normally measured by the training loss function. It indicates the gap between the learned values and the labeled data. Convergency is another important metric. The time in convergency should be as low as possible. Other metrics such as communication cost, energy efficiency of the edge devices, reliability and data privacy are possibly concerned by different stakeholder.

\subsection{Challenges of Model Learning in Edge AI}

Model learning in edge computing pertains to the technical area of Distributed Machine Learning (DML). Compared with existing DML in cloud, model learning on edge faces a few challenges \cite{ref_1} \cite{add_15}. First, the data sources in edge computing are generated in real time from the edge and/or end devices. The data samples on each device are usually non-iid data. Existing DML in cloud has the data source pre-stored in a centralized cloud storage. The data sources are allocated to the workers/servers for data-parallel processing such that the data samples at each worker are IID data. Second, the computing environment to perform the learning task in Edge AI are challenging. The devices have heterogeneous compute capabilities and are connected by bandwidth-limited and intermittent wireless networks. While in existing DML on cloud, the learning tasks are done in a cloud cluster where the servers have powerful capability and the network connecting the servers has guaranteed and stable network bandwidth. The computing environment makes it difficult for a large number of devices to synchronize in the learning process. Third, model learning for Edge AI has a complex trustable environment. In Edge AI, the learning task allows limited data exchange among the edge devices due to privacy concern of the data owners, while in cloud learning task, data frequently exchanges among the workers by particular data shuffling algorithms due to data co-existence in the same trustable environment.

\subsection{Motivations of Proposing E-Tree}

To solve the above challenges, substantial research works have been proposed for model learning in Edge AI, which can be classified into three categaries, i.e., \textbf {centralized approaches}, \textbf{fully distributed approaches} and \textbf{decentralized approaches}. Representative centralized approaches are federated learning \cite{ref_16}\cite{add_9}, where a central master is to aggregate the model parameters from multiple workers/slavers and then updates the aggregated model back to the workers. The approach has several drawbacks. The worker with very slow speed in local model updating can be a straggler. Stragglers of some workers affect significantly the convergence speed. Besides, the constrained communication bandwidth from all the workers into the central master may become the performance bottleneck. Also the central master faces single point of failure. Fully distributed approach like Gossip learning \cite{ref_8} allows the model aggregation on every edge device. The edge device does the local updating and sends its model randomly to nearby nodes. Nearby nodes receive the model and aggregate it with its own model, and then sends it again to the neighboring nodes. The model aggregation is done asynchronously over all the edge devices. The fully distributed approach avoids the single point of failure and performance bottleneck caused by central master, but it neglects the data distribution due to the randomness in model aggregation and thus yields slow convergence speed and model inaccuracy.

The recently proposed Hierarchial Federated Learning (HFL) pertains to a decentralized approach \cite{add_10}. HFL typically adopts a three-layer aggregation structure by adding a middle layer model aggregation into the original federated learning. HFL is supposed to be built on a tiered infrastructure, and the layers of the aggregation structure is inherently determined by the number of infrastructure tiers. Therefore, HFL has a static aggregation structure and is not suitable for a dynamic mesh network with mobile edge devices such as multi-robot systems, vehicular network and drone network. In the mesh network, the aggregation structure needs to be configurable on demand according to the dynamic network.

Considering the limitations of existing approaches, we want to step ahead by proposing E-Tree learning. E-Tree is a \textbf{configurable decentralized approach}. It uses a well-designed tree structure for localized and hierarchical model aggregation. E-Tree is suitable for any infrastructure such as mesh networks. The structure of the aggregation tree including the number of layers and node grouping is dynamically built according to network topology and data distribution.

\section{E-Tree Learning Framework Design}

\subsection{Structure Overview}
\begin{figure}
  \centering
  \includegraphics[width=3.5in]{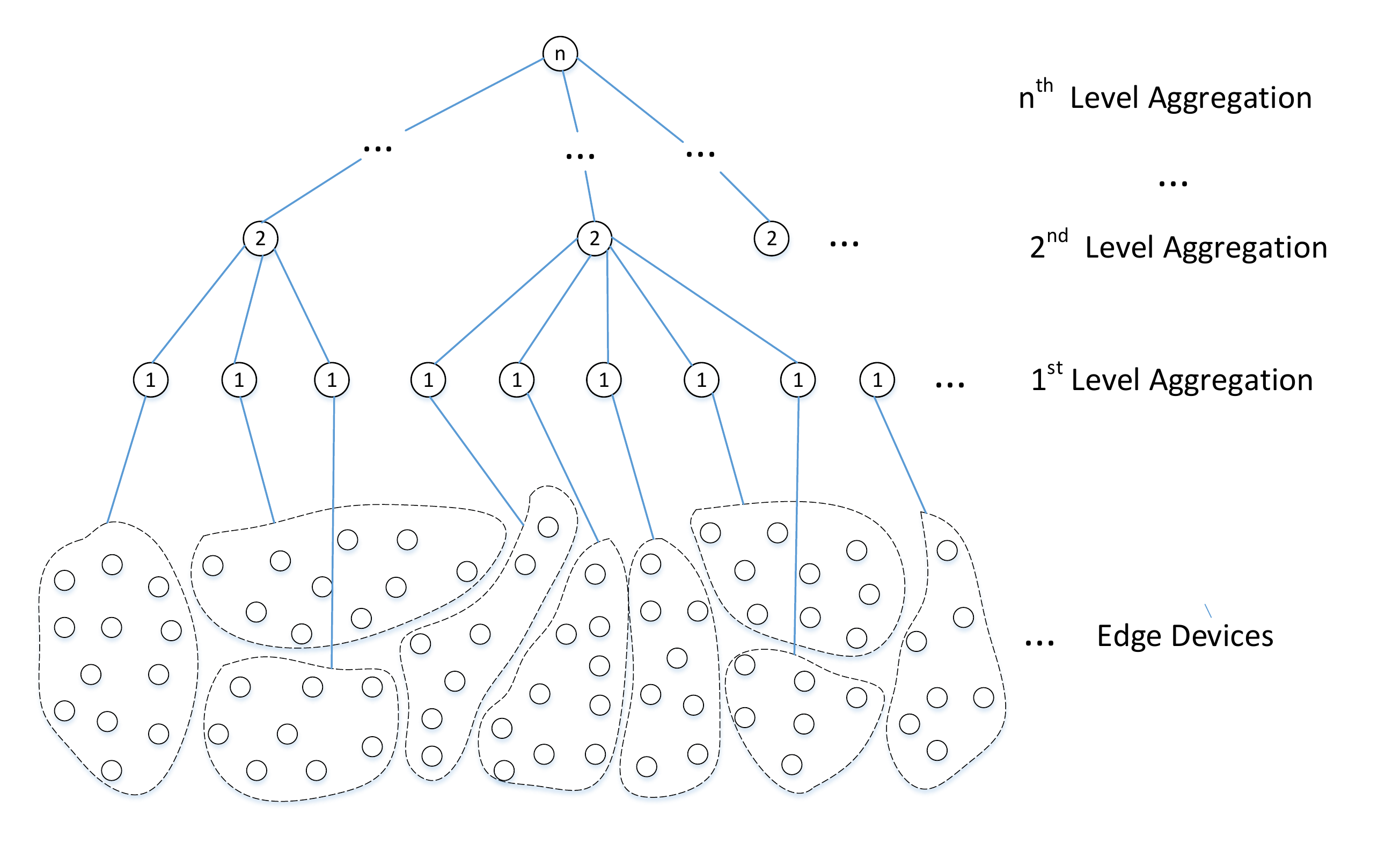}\\
  \vspace{-0.3cm}
  \caption{Structure Overview of E-Tree learning}\label{structure}
\end{figure}

Fig.\ref{structure} shows an overview of E-Tree learning structure. It consists a hierarchical tree based structure for the model aggregation. In the structure, the leaf-nodes at the bottom layer represent the edge devices involved in the learning. The non-leaf nodes represent model aggregation. We name the non-leaf nodes by {\itshape aggregation nodes}. The model aggregation follows a bottom-up approach. The edge devices are firstly grouped for the 1st level (bottom level) aggregation. The grouping is done according to the distribution of data owned by the edge devices as well as the network distance. Similar to existing federate learning, E-Tree learning also requires an initial model to start the training. Within a group, the edge device generates a model update using its own dataset. The model updates are aggregated in the group. To do in-group aggregation, an aggregation/routing tree is normally constructed to save the communication cost and reduce the latency. It surely supports the aggregation on a centralized device.

After the 1st level aggregation, E-Tree learning further aggregates the model updates from the groups. The inter-group model aggregation uses the same idea with the in-group aggregation. The aggregation is recursively done level by level from bottom to up. The top level aggregation is the root of E-Tree learning. It aggregates the model updates and then sends back the result to all the edge devices in the same routing paths with the aggregation. The edge devices receive the results and updates the local model again. Then, E-Tree learning begins the next-iteration model aggregation.

\subsection{Building E-Tree}

\begin{figure*}[!t]
\centerline{
\subfloat[physical network topology]{\includegraphics[width=2.0in]{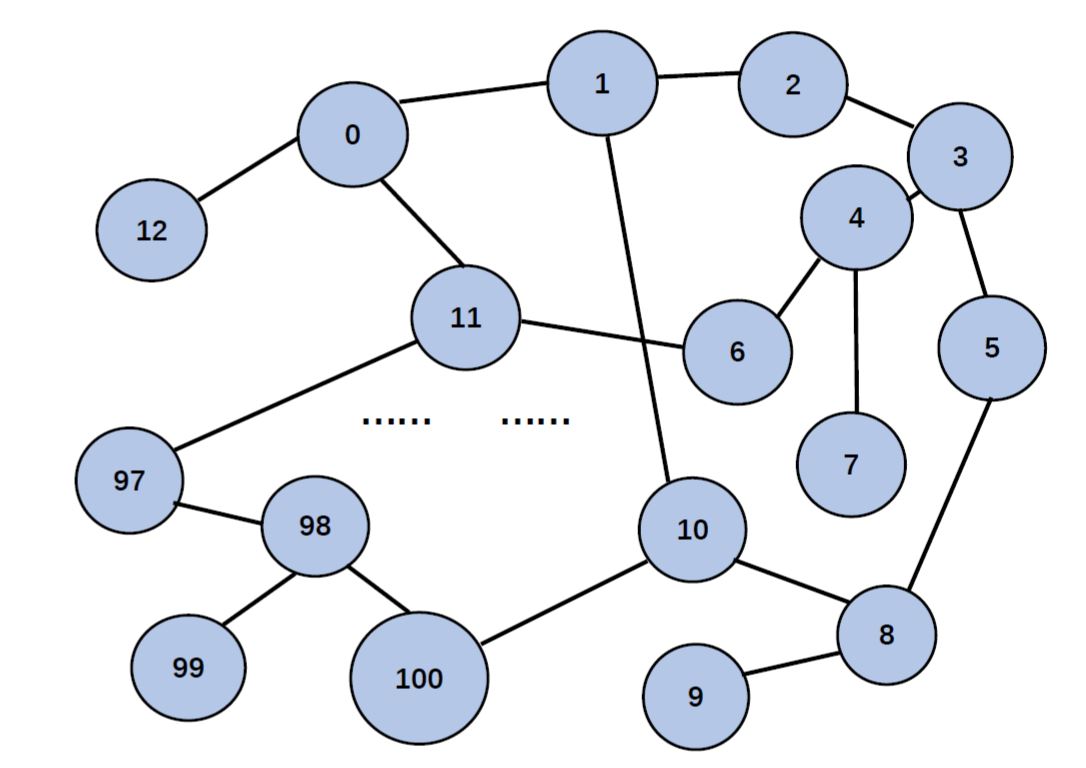}\label{topoGraph}}
\hfil
\subfloat[E-Tree aggregation topology]{\includegraphics[width=3.0in]{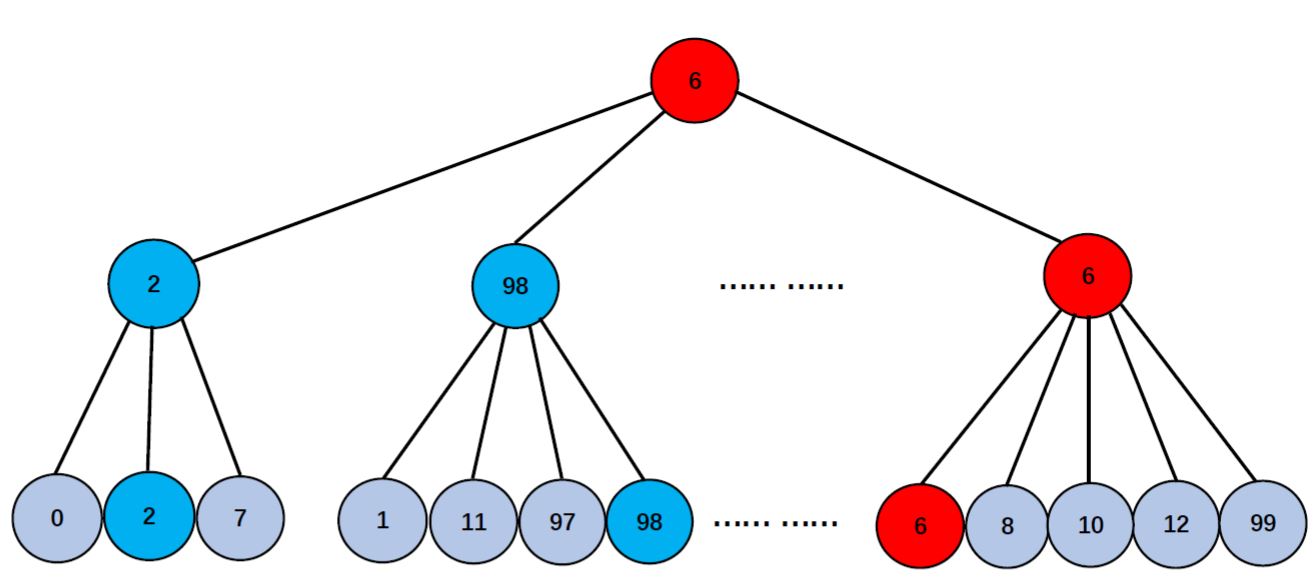}\label{etree_3l}}
}
\label{fig:count}
\caption[]{An example of building a three-layer E-Tree on a physical network topology}
\vspace{-0.4cm}
\end{figure*}

In this section, we introduce how to build E-Tree based on a general network topology. We abstract the underlying physical network including all the edge devices as a graph, in which the nodes represent the edge devices and the link represents the network connections among the edge devices. The link in the graph has a weight denoting the transmission delay of the model parameters. Fig.\ref{topoGraph} shows an example of network topology graph. It should be noted that the graph is not necessarily fully connected. The graph represents the physical topology, E-Tree structure built on top of the physical network topology is a logic topology representing which devices are aggregated together and at which device the aggregation is done. Fig.\ref{etree_3l} shows the aggregation topology with a three-layer E-Tree on top of the physical network.

\begin{table}[t]
\renewcommand{\arraystretch}{1.1}
\caption{Symbols and notations}
\begin{center}
\begin{tabular}{|c|p{7cm}|} %l(left)居左显示 r(right)居右显示 c居中显示
\hline
\textbf{Symbol} & \textbf{Description}\\
\hline
$N$ & The number of edge devices in the physical network; \\
$d_{i,j}$ & The network distance measured by transmission delay among the edge devices $i$ and $j$; \\
$l$ & The number of layers in a E-Tree structure, $l=1, 2, 3, \cdots$, where $l=1$ represents the bottom layer of leaf nodes; \\
$N_l$ & The number of nodes at the $l$-th layer in E-Tree; \\
$K_l$ & The number of clusters/groups at the $l$-th layer in E-Tree; \\
$a_l$ & The aggregation frequency of nodes at the $l$-th layer; \\
$\delta$ & A threshold parameter of the KMA algorithm to constrain the difference of the clusters in average model accuracy; \\
\hline
\end{tabular}
\label{tab1}
\end{center}
\end{table}

Supposed that there are $N$ nodes in the physical network $G$, and the node IDs is denoted as $\{0,1,2,...,N-1\}$. The transmission delay between any two nodes $i$ and $j$ is denoted as $d_{i,j}$. E-Tree is built from the bottom to top. We denote the layers of E-Tree as $l$, and the first layer is the bottom layer. First, all nodes in the physical network are served as leaf nodes of E-Tree. We denote the leaf nodes at the bottom layer of E-Tree as $N_1$. Then we use node clustering algorithms, which are detailed in Section~\ref{nodeClustering}, to divide the nodes of $N_1$ into $K_1$ groups, and denote the nodes in each cluster as $\{C_{1,1}, C_{2,1}, ..., C_{K_1,1}\}$. We use $k_1$ to denote the index of the clusters in the first layer. Next, we find the center node $n_{k_1, 1}$ of each cluster $C_{k_1,1}$ by

\begin{equation}
n_{k_1, 1} = \mathop{\arg\min}_{i \in C_{k_1, 1}} \sum_{j \in C_{k_1,1}, j \neq i} d_{i,j}.
\label{eq1}
\end{equation}

After the center nodes of these clusters are found, we use these center nodes as aggregation nodes at the second layer in E-Tree, which are denoted as $N_2$. Each node of $N_2$ is connected to all of the nodes in the corresponding cluster. Finally, we find the center node of $N_2$ in the same way, and use it as the root node of a 3-layer E-Tree. The root node is connected with all of the nodes in $N_2$. As shown in Fig.~\ref{etree_3l}, the nodes in dark color are center nodes.

\subsection{Model Training and Aggregation}

The leaf node of E-Tree is responsible for training a local model using its own data samples, while an internal node aggregates the models from its children and updates an aggregated model back to the children. Note that the nodes which are chosen to be the center nodes finish model training and aggregation on the same device.

Specifically, we first initialize a model for the root, and the root sends the model to all of the nodes in $N_2$. After nodes in $N_2$ receive the model, they save the model locally and send it to their children without any processing. Nodes in $N_1$ then receive the model and begin training on it with their local data using Stochastic Gradient Descent (SGD). After the training is finished, the node computes the updates of the model and sends the updates to its parent. After the parent in $N_2$ receives all the updates from each child, it starts the model aggregation. The aggregation is done by computing the average of all the updates and adding it to the current model in the parent. Then the parent sends the updated model to its children for training. The bottom-up aggregation is recursively done level by level.

We define the {\itshape aggregation frequency} of the layer $l$ as $a_l$. It means the aggregation node at layer $l$ does one global aggregation to its parent every $a_l$ times of local aggregation from its children. For instance, nodes in $N_2$ send their models to their parents after finishing $a_2$-$th$ aggregation from the children. Similar to the nodes in $N_2$, the root starts the aggregation after receiving all the models from its children. After the aggregation on the root is done, the root sends its new model downwards and the new turn is processed as mentioned above.

\subsection{Extension of E-Tree to Multiple Layers} \label{extension}

We show the E-Tree structure with 3 layers as an example above. However, a network topology graph may contain thousands of nodes which might have long transmission delay with each other. We can further extend E-Tree into more layers in order to fit a larger scale network.

Transmission delay between any two nodes in the same cluster should be relatively short to make sure there is no straggler. In order to achieve this purpose, the number of clusters in layer 1, i.e., $K_1$, should be increased, which means the number of nodes in $N_2$ is increased. Therefore, we can further divide $N_2$ into $K_2$ groups, and the center nodes of these groups form layer 3, denoted as $N_3$. We define the aggregation frequency of layer 3 as $a_3$. Finally, the center node of $N_3$ becomes the root of 4-layer E-Tree. The process of model training and aggregation in a 4-layer E-Tree is the same as that in 3-layer E-Tree.

\section{E-Tree Refinements}

Next, we discuss concrete issues in the design of E-Tree Learning. First, we present how to optimize the structure of E-Tree learning in terms of nodes clustering. Second, we solve the synchronization problem by controlling the aggregation frequency of each aggregation node. Third, we discuss how to schedule the aggregations onto the physical edge devices. Last, we introduce the issue and our solution arising from the Non-IID data.

\subsection{Device Clustering} \label{nodeClustering}

As shown in Fig.\ref{structure}, E-Tree learning adopts a tree based hierarchical structure to aggregate the model parameters. We consider the structure optimization of E-Tree learning, i.e., by answering the questions how many aggregation nodes each level of E-Tree learning should have. To optimize the structure, E-Tree learning clusters the edge devices before the 1st level of aggregation. It determines which edge devices should be grouped together for model aggregation. The clustering depends on several factors including data distribution on the edge devices, network topology and resources owned by the edge devices. Existing network clustering algorithms group the devices according to the physical distance in the network. The devices with low communication cost among each other are grouped together. However, in E-Tree learning, data distribution is another important factor to influence the clustering. Meanwhile, the clustering guarantees load balancing among the clusters. The data set involved in each cluster are balanced in data mount. The resources in computation and energy of the edge devices are also taken into account for load balancing.
%Second, we describe the issue to optimize the number of levels in a E-Tree learning structure. The centralized model aggregation in federate learning can be considered as a special case of E-Tree learning with one level of aggregation, at which there exists only one aggregation node. One opposite method to federate learning is sequential aggregation which sequentially traverses the edge devices. During each visit of edge device, the model updates is aggregated with the local model on the device. The sequential aggregation can also be considered as a special case of E-Tree learning, where the number of levels is equal to the number of edge devices and there exists only one aggregation node at each level. Obviously sequential aggregation fully utilizes the capability of each edge device, but it is slow as each round of aggregation needs to visit all the devices. In E-Tree learning, the number of levels is decided by the scale of the edge devices and networks. More levels are suitable for a large scale of network. However, too many levels would lead to long latency in the level by level model aggregation.

\subsubsection{The Influence of Data distribution to Device Clustering}

The device clustering algorithm should make sure that the transmission delay is relatively short between each node and the center node, such that there is few stragglers. Besides transmission delay, we also study how data distribution of nodes influences node clustering. In reality, the data distribution is often non-IID. For example, each node might only have data of few classes or even only one class. Such non-IID data leads to low accuracy and slow convergence. If each node owns samples of more classes, the accuracy can be improved \cite{ref_5}. We want to know how the data distribution affects the node clustering via experimental studies. In our experiments, the data distribution is set to non-IID, where each node only owns samples of 1 or 2 classes.

\begin{figure}[t]
	\centering
	\subfloat[$K_1 = 5$]{
	\includegraphics[scale=0.194]{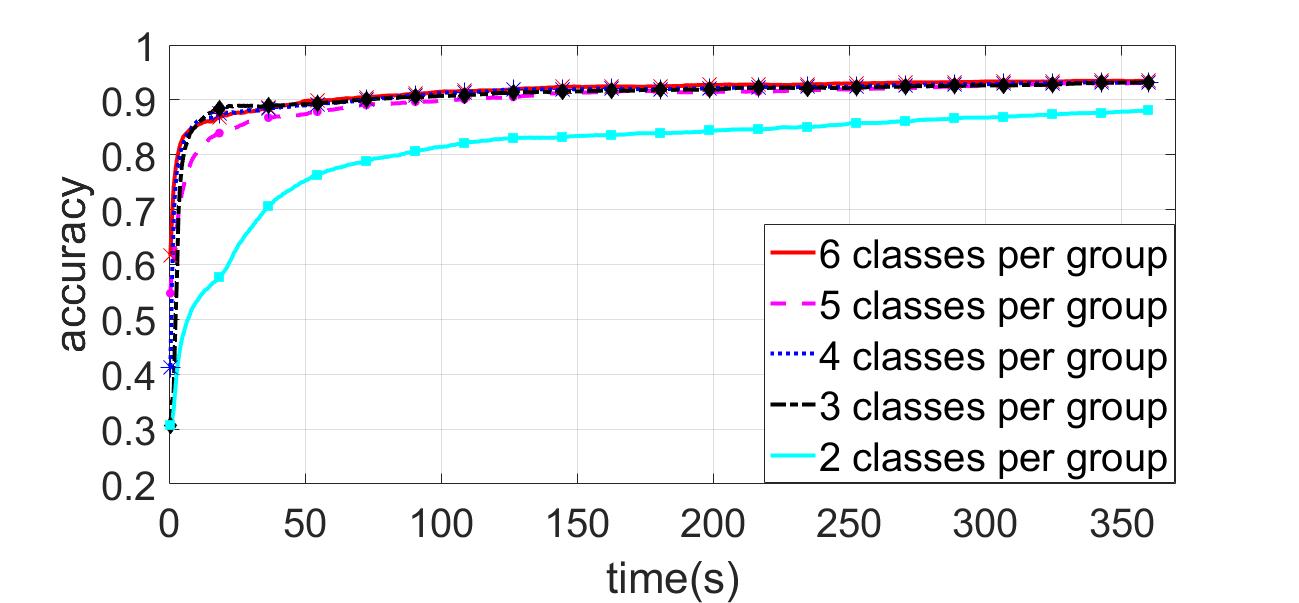}}
	\label{la}
	\subfloat[$K_1 = 8$]{
		\includegraphics[scale=0.194]{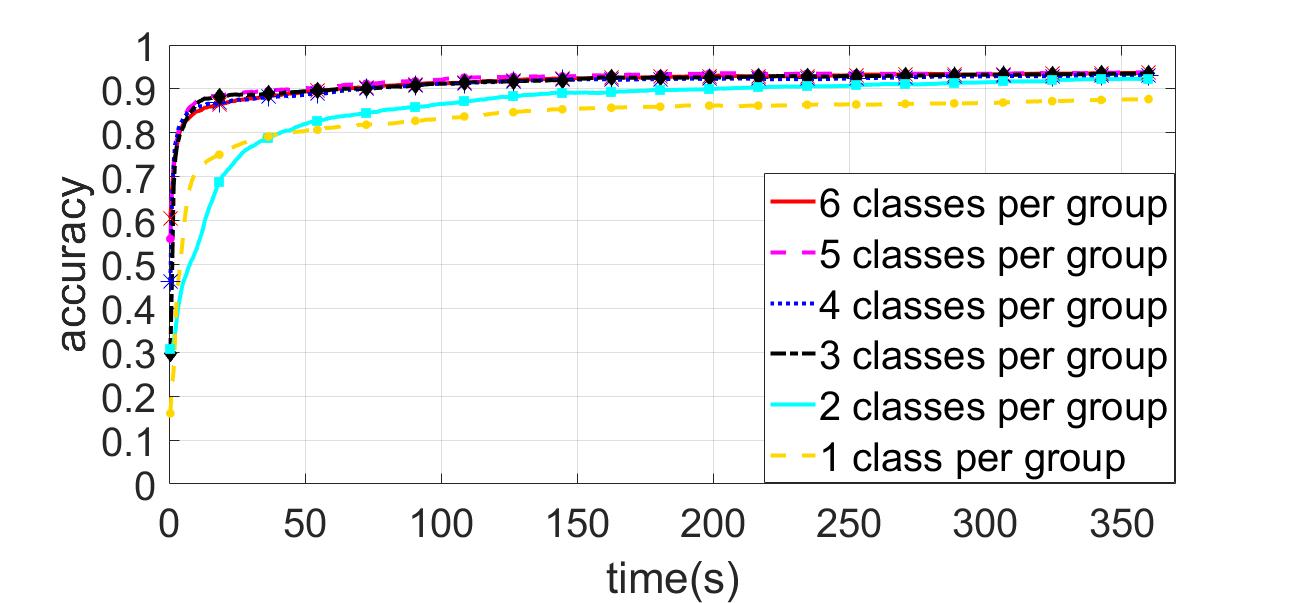}}
	\label{lb}
	
	\caption{Comparing results of different number of classes each group owns.}
	\label{diff_num_of_classes}
\end{figure}

To further study the influence made by data distribution in E-Tree, we perform several experiments to compare the model accuracy of different number of classes each group of a 3-layer E-Tree owns. In order to omit the influence made by transmission delay, we use a fully connected network topology graph with 100 nodes where transmission delay between any two nodes is the same. The dataset is HAR and the parameters are detailed in Section~\ref{sec_eval}. The experiment results are shown in Fig.~\ref{diff_num_of_classes}.

As the result shows, when there is only 1 or 2 classes per group, model accuracy and convergence rate are obviously decreased. When there are 3 or more classes per group, model accuracy and convergence rate are about the same and better. Hence, if the node clustering algorithm guarantees there are as many classes per group as possible, the final model accuracy and convergence rate can be improved. This is the main motivation of our proposed clustering algorithm discussed in the following.
Next, we introduce our proposed node clustering algorithm named by KMA, and then briefly introduce the baseline clustering algorithms as a comparison to KMA.

\subsubsection{Node Clustering Algorithm based on K-Means and Average Accuracy (KMA)}

K-Means clustering algorithm takes into account transmission delay while omitting data distribution. As shown before, class distribution among groups also influences the final model accuracy. Therefore, in this section, we design an algorithm based on K-Means that considers both transmission delay and class distribution, namely KMA. For each node, KMA has a common test set to compute the average accuracy of the model that is pre-trained on the local data of each node. If the nodes have similar class distribution, the pre-trained model accuracy of the nodes are also similar. So we can estimate the similarity of the nodes' class distribution through the pre-trained accuracy. KMA clusters the nodes based on both transmission delay and pre-trained accuracy, and limits the difference between the average pre-trained accuracy of the nodes in a cluster and the average pre-trained accuracy of all nodes in the graph. We define a predefined threshold $\delta$ to limit the difference. The process is shown in Algorithm 2.

\SetAlFnt{\small}
\IncMargin{0.5em}
\begin{algorithm}
\SetKwInOut{Input}{Input}
\SetKwInOut{Output}{Output}
\Input{the number of clusters $K$, network topology graph $G$, nodes to be clustered $\{n_1, n_2, ..., n_N\}$, $N > K$, test set $D_{tes}$, allowed difference $\delta$, number of pre-training rounds $r$}
\Output{clustering result $\{C_1, C_2, ..., C_K\}$}
Compute the minimum transmission delay matrix $[d_{i,j}]$ based on graph $G$\;
On each node in the graph, train an initial model on its local data for $r$ rounds\;
Use $D_{tes}$ to test the pre-trained models of the nodes and compute the accuracy $acc_j$ of each node $j$\;
Compute the average pre-trained accuracy of all nodes in the graph $acc_{avg} = \frac{1}{N} \sum_{j=1}^{N} acc_j$\;
Randomly select $K$ nodes as initial center nodes, and initialize a cluster for each center node\;
\For{each node $n_i$ to be clustered} {
    Sort the current $K$ center nodes according to their distance to $n_i$ in an ascending order\;
    Take the first half of the sorted center nodes, and denotes the corresponding clusters by $C_1, C_2, \cdots, C_{[K/2]}$ \;
    \For{each cluster $C_{j}$, where $1 \leq j \leq [K/2]$}{
        Add $n_i$ to the cluster $C_j$ and compute the average pre-trained accuracy $acc_k$ of the nodes in the cluster\;
        \If{$|acc_k - acc_{avg}| < \delta$}{
            Assign $n_i$ to the cluster $C_{j}$ and goes to Line 17\;}
        \Else{
            Remove $n_i$ from the cluster $C_j$\;}
    }
    \If{$n_i$ has not been assigned to a cluster successfully} {
        Assign $n_i$ to the cluster whose center node has the shortest transmission delay to it \;
    }
    Update the center node of the clusters using Equation \eqref{eq1}\;
}
Repeat line 6-17 until the center node of clusters does not change\;
\Return
\caption{KMA Algorithm}\label{algorithm_2}
\end{algorithm}

To highlight the difference of pre-trained accuracy among nodes, $D_{tes}$ is sampled from the original test set and the number of samples of each classes is different in $D_{tes}$. KMA makes the difference between the average pre-trained accuracy of the nodes in a cluster and the average pre-trained accuracy of all the nodes as small as possible. It guarantees that the nodes with similar class distribution are not always gathered together to significantly increase/decrease the average accuracy of a cluster. Hence, each cluster can obtain data samples with as many classes as possible. The complexity of Algorithm 2 is $O(N\times K)$, where $N$ is the number of nodes and $K$ is the number of clusters.

\subsubsection{Benchmark Clustering Algorithm}

We introduce two naive clustering algorithms as the benchmarks to compare with our proposed KMA algorithm. The first benchmark is the {\itshape K-Means} clustering algorithm without considering the data distribution on the nodes. We partition the nodes into $K$ clusters based on the network transmission delay. K-Means can help us to assign the node to the cluster where transmission delay is the shortest between them. Note that in this algorithm, we only consider transmission delay. The second benchmark is named by {\itshape Ununiform KMA}. It generates the clusters such that the class distribution among the clusters is ununiform. Each cluster owns data samples with few classes. The difference between the average pre-trained accuracy of the nodes in a cluster and the average pre-trained accuracy of all nodes in the graph is relatively great.

\subsection{Controlling Aggregation Frequency}

For an aggregation node, it aggregates model updates from the children nodes and then sends the result to its parent node. The model updates would not be sent back to the edge devices until the model is aggregated to the root node in E-Tree learning. In this case, the aggregation frequency of all the aggregation nodes are the same, which is named by {\itshape strong synchronization}. The aggregation that finishes at an early time must wait for the other aggregations at the same level. This approach does take into account the heterogeneity of edge devices. Also the workloads in aggregation are possibly different.

E-Tree learning adopts a {\itshape week synchronization} method. It allows different aggregation frequency at various aggregation nodes in order to fully utilize the resources of the heterogeneous edge devices. The aggregation frequency at the aggregation nodes depends on the workloads and the resources. E-Tree learning constrains that the frequency of an aggregation node should be integer times of the frequency of its parent node. The frequency of a parent aggregation node should be a common multiplier of the frequency of all its children nodes. With this constraint, E-Tree learning has a simple heuristic to determine the aggregation frequency at each aggregation node. The objective is to minimize the loss function given the limited resources of the edge devices. The decision depends on the data distribution, network dynamics and model characteristics. In weak synchronization, when the root node finishes a global aggregation, the results are transmitted back to the leaf nodes (edge devices). We say an iteration of model aggregation is completed. Then, E-Tree learning begins the next iteration.

In order to deal with the network dynamics, E-Tree learning enables the change of structure in the next iteration of aggregation. Thus, E-Tree learning has a planner to dynamically adjust the structure of the aggregation tree. In future work, we will consider to apply the reinforcement learning augmented with a Graph Neural Network (GNN) to determine the structure online. GNN is used to represent the dynamic characteristic of the underlying edge networks including the features of edge devices and communication links. With these input features, reinforcement learning learns the tree structure.

\subsection{Scheduling Model Aggregation}

In E-Tree learning, an aggregation node represents an operation to aggregate all the model updates from the children nodes. Typical example of the aggregation function is the averaging function. Given an aggregation node in E-Tree learning and the places of its children nodes, we need to determine the place of the aggregation node and as well build an aggregation tree to transmit the model updates from the children nodes to the aggregation node. This aggregation tree is actually a physical routing tree, which is different with the structure of E-Tree learning shown in Fig.\ref{structure}. To distinguish it from the tree based structure in E-Tree learning, we name it as a {\itshape routing tree}, in which the node represents an aggregation operation done at an edge device, and the arc represents a communication link in the network. We can imagine that every aggregation node of E-Tree learning in Fig.\ref{structure} has a routing tree to physically connect to its children nodes.

The scheduling problem needs to map the logic nodes of the routing tree onto the physical devices, and also determine the execution order of these logic nodes. The objective is to minimize the latency in the aggregation and the communication cost over the network. If we consider a single aggregation node of E-Tree learning, and assume that its children nodes have been allocated to fixed devices, this scheduling problem is not different from existing aggregation problem in sensor networks \cite{add_8}. However, as E-Tree learning includes multiple levels of aggregation, an aggregation node acts as a root in the routing tree connecting to its children nodes, while it also acts as a leaf node in the routing tree connecting to its parent node. Thus, we could not independently build the routing tree for each aggregation node in E-Tree learning. On the contrary, we need to optimally build a global routing tree for all the aggregation nodes in E-Tree learning. The scheduling problem for the global routing tree is new and more challenging than existing aggregation problem in sensor networks. To schedule the global routing tree onto the physical edge devices, we take into account of the structure of E-Tree learning, the resources of the devices, the resources of the communication links and the data distribution.

\textbf{Reliability Issue in Scheduling}. Due to frequent device and link failures in edge computing network, high reliability is an important goal of the E-Tree design. Considering the reliability, the scheduling problem is to allocate the logical aggregation operations onto the physical devices, with the objective of minimizing the time of a global aggregation at the root node while satisfying the reliability requirement. To model the reliability, we first present the fault model. We consider both computation failure of the model aggregation and the transmission failure of the model parameters. The failure of model aggregation can be caused by the hardware/software crash of the physical device where the model aggregation is performed. It can be denoted by a constant probability. The transmission failure is caused by the link failure of the network. Assume that all the links have the same failure probability. The transmission failure of the model parameter along a path increases depending on the number of hops of the path.

To satisfy high reliability requirement, we replicate the model aggregation on multiple physical devices. Meanwhile, we leverage the re-transmission mechanism to increase the transmission reliability. Then, the schedule problem needs to determine the number of replicas of each aggregation operation, and the number of retransmission among the aggregations in E-Tree. This is a challenging problem which needs to balance the reliability and completion time under constraint of resources.

Existing heuristics have been proposed to schedule DAGs (Directed Acyclic Graph) with task replication onto the heterogeneous machines \cite{add_17} \cite{add_18}. The schedule problem in E-Tree can be transferred into the same problem by abstracting the logical aggregation operations as dependent tasks. The tree based task graph is a special instance in the DAG model. The difference between this problem and the existing DAGs scheduling is that the model transmission for the aggregations can also fail along the network path. The reliability of an aggregation depends on the success probability of the model transmission as well as the aggregation operation itself. Thus, given an E-Tree structure and the allocation of replicas of aggregations, we can calculate the reliability of the root node by using a bottom-up recursive approach. The reliability of low-layer aggregation is calculated first and then are used to update the reliability of up-layer aggregation. It is done recursively until the reliability of the root node is obtained. In our future work, we will further model this problem and develop solutions for reliable scheduling of aggregations with consider the link failures. The solution is to find near-optimal replication and scheduling of aggregations under the reliability requirement, aiming to minimize the aggregation time in a training round.

\subsection{Incorporating Non-IID Data}

\begin{figure}
  \centering
  \includegraphics[width=3.2in]{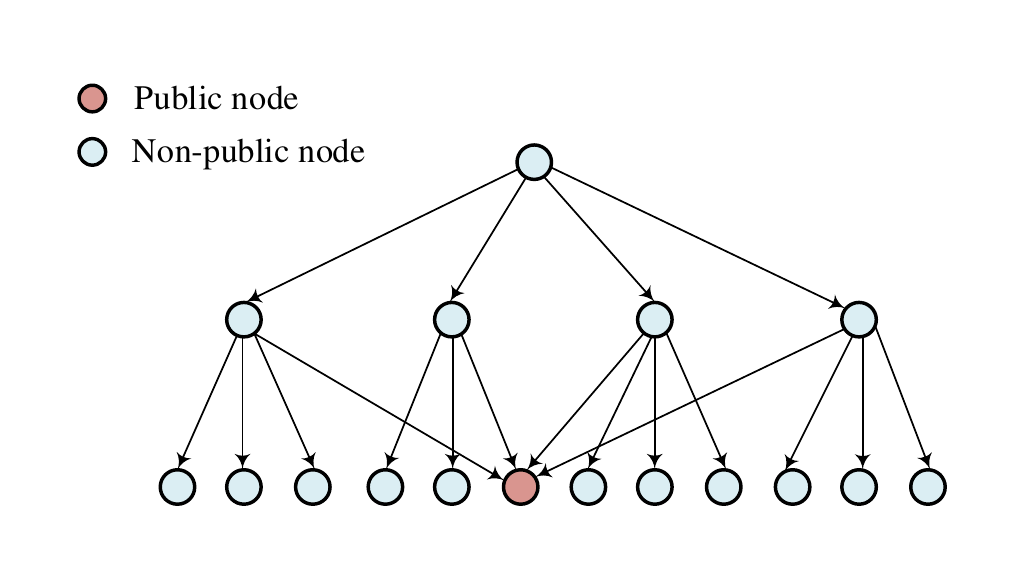}\\
  \vspace{-0.3cm}
  \caption{Structure refinement for solving Non-IID data}\label{refinement}
\end{figure}

Non-Independently and Identically Distribution (IID) data on the edge devices greatly impact the loss function of the model learning. In order to solve the problem, we allow the overlap of clusters at each level of E-Tree learning. At the 1st level aggregation, an edge device can participate in multiple clusters. The purpose is to allow the clusters to have a common shared data, which has been demonstrated to reduce the loss function in training a model \cite{ref_5}. At the other levels of E-Tree learning, an aggregation node can have multiple parent nodes. Fig.\ref{refinement} shows the refined structure of E-Tree learning for solving the Non-IID data distribution. The larger the overlap among the cluster is, the less loss function the model learning has. Meanwhile, the communication cost would be greater. Thus, E-Tree learning optimizes the overlapping in order to balance a trade-off among the communication cost and loss function.

Considering the 1st level aggregation in E-Tree learning, we name the edge device which has multiple parents as the {\itshape public node}. It is required that a public node should participate in all the aggregation nodes in the upper level. In this way, the data of the public node can be considered as a common shared dataset in the upper level aggregation. The shared dataset can help reduce the loss function. How many and which edge devices should be selected as the public nodes? In term of network resources, the edge devices located at the centroid of network would be selected to minimize the transmission cost. The data amount and distribution also impact the problem decision. We model the public node selection problem and design efficient heuristics to minimize the loss function under the constraints of resources.

Algorithm \ref{algorithm_4} presents the pseudo-code of the public node selection algorithm. Assuming that the nodes have been clustered using the KMA algorithm in Section \ref{nodeClustering}, we probe each node which has a low average distance to all the $K$ cluster centers. When a new node is probed as the public node, we check whether the deviation of the data distribution among the clusters is still within the required threshold. If it is, the node is then selected as public node; otherwise, the algorithm probes the next node. As we demonstrate by experiments in Section \ref{nodeClustering}, if the clusters have similar data distribution among each other, then E-Tree converges better model accuracy. Thus, we leverage the deviation in data distribution among the clusters to detect whether a node should be selected as public node. The algorithm has a parameter $\gamma$ to control the number of the public nodes. The nodes with the $\gamma \times N$ nearest distance to all the clusters are considered as the candidate nodes which our algorithm probes. The complexity of this algorithm is $O(N \times K)$.

\SetAlFnt{\small}
\IncMargin{0.5em}
\begin{algorithm}
\SetKwInOut{Input}{Input}
\SetKwInOut{Output}{Output}
\Input{network topology graph $G$, clustering result $\{C_1, C_2, ..., C_K\}$, the accuracy of the pr-trained model on each node $acc_j$, $1<j<N$, the threshold $\gamma$ to control the number of selected nodes, $0 < \gamma < 1$}
\Output{the set of public nodes $\mathcal{S}_{pub}$ }
Compute the average pre-trained accuracy of all the nodes $acc_{avg} = \frac{1}{N} \sum_{j=1}^{N} acc_j$\;
Compute the average pre-trained accuracy of the nodes within each cluster $acc_k$, where $1 \leq k \leq K$\;
Compute the distance of each node $n_j$ to the other clusters $r_j = \frac{1}{K}\sum_{i=1}^{K} Dist(n_j, C_i)$\;
Sort the nodes according to an ascending order by $r_j$ and select the first $\gamma \times N$ nodes as the candidate nodes\;
\For{each node $n_j$ from the candidate nodes} {
    Add the node $n_j$ into the other $K-1$ clusters in which the node $n_j$ does not exist\;
    Update the average accuracy $acc_k$ of each cluster\;
    \If{$\forall k \in [1, K]$,$|acc_k - acc_{avg}| < \delta$}{
        Add $n_j$ to the set $\mathcal{S}_{pub}$ of public nodes\;}
    }
\Return $\mathcal{S}_{pub}$
\caption{Public Node Selection Algorithm}\label{algorithm_4}
\end{algorithm}

Although structure refinement is an effective method to reduce the loss function in model learning from the Non-IID data, we also adopt {\itshape data redistribution} (or shuffling) as an complementary method. This method redistributes some of the data among the edge devices before calculating the model update in each iteration. Due to high communication overhead in data distribution over all the edge devices, we adopt the coding mechanism to reduce the communication overhead. The basic idea is to package some of the data samples at the sender, and send the package to multiple receiver. The receiver can decode its desired data from the package using the data stored at the receiver. Based the coding mechanism, E-Tree learning determines which data samples should be packaged together and schedule the package transmission to minimize the communication cost.

%We have developed a coding based distributed data shuffling method to solve the problem in our previous work. It is applied in E-Tree learning to deduce the communication cost by data shuffling.
\section{Evaluations}\label{sec_eval}

\subsection{Evaluation Setup}
We have done simulations to compare the performance of E-Tree learning with federated learning and Gossip learning. In the simulations, we develop E-Tree learning based on an open source benchmark framework which was used for comparing federated learning and Gossip learning \cite{ref_8}. As E-Tree learning is designed for Edge AI, the simulation environment is set up to incorporate the features of edge computing. Details of the experiment setup are as follows.

We generate a network with 100 edge devices and 300 links connecting the devices. The topology is randomly generated. As the links have different bandwidth resources, the latency of transmitting model parameters on the links are different, which yield a uniform distribution with a mean value of 50 ms and a standard variance of 50 ms. We use {\itshape Human Activity Recognition (HAR) Using Smartphones Dataset} from the UCI machine learning repository in our simulation \cite{add_11}. The dataset contains 10299 samples, where 7352 samples are distributed on the edge devices for training, and 2947 samples are used for testing. The dataset has 6 classes, and each data sample has 561 features. We choose the model of {\itshape Softmax Regression} to classify the human activities.

As we want to evaluate the performance of E-Tree under different data distributions, we respectively generate two data distributions according to the classes/labels of the data samples, i.e., IID and Non-IID. Under IID training data, the data samples with the same class are uniformly distributed onto the edge devices, and thus every edge device has all the 6 classes of data samples. Under Non-IID training data, every edge device has 4 classes of data samples which are randomly selected out of the 6 classes. Table \ref{environment_para} shows the parameters of the simulation setting. The detailed configurations of the three methods are as follows.

\begin{table}[!t]
\renewcommand{\arraystretch}{1.1}
\centering
\caption{Parameters of the simulation environment}\label{environment_para}
\begin{tabular}{p{5.1cm}|l}
\hline\
 Parameters & Values \\
\hline
The number of edge devices & 100 \\
Number of links among the edge devices & 300 \\
Transmission delay of links with mean and standard variance & 50 ms, 50 ms \\
Size of training dataset & 7352 \\
Size of testing dataset & 2947 \\
No. of classes & 6 \\
No. of features & 561 \\
No. of data samples per edge device & 73 \\
Learning model & Softmax Regression \\
Learning rate & 0.02 \\
Percentage of client selection $C$ & 1 \\
Simulation time & 30 $\times$ 1000 ms \\
\hline
\end{tabular}
\end{table}

\begin{figure*}[!t]
\centerline{
\subfloat[Convergence in terms of round under IID data]{\includegraphics[width=3.0in]{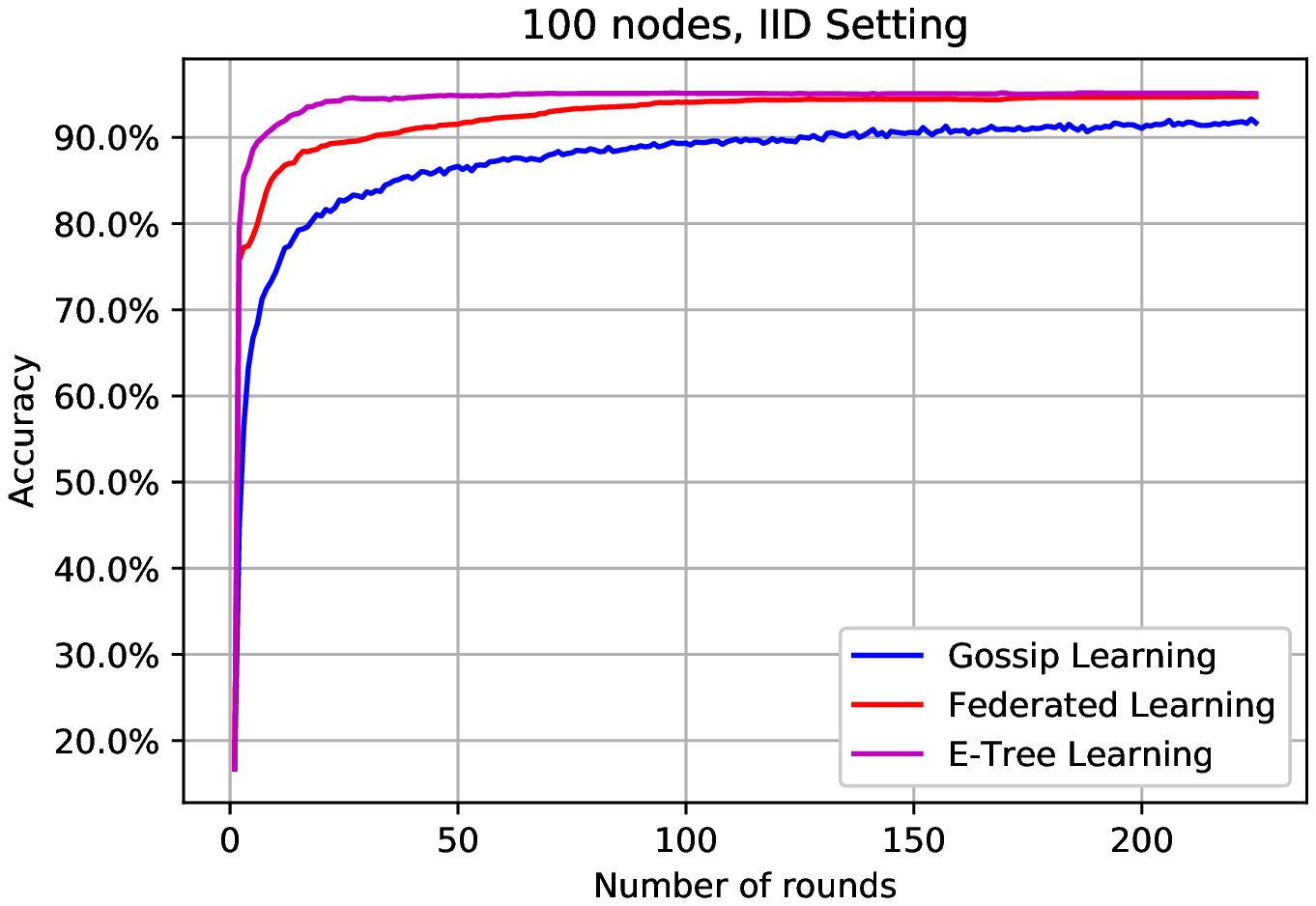}\label{conv_round_iid}}
\hfil
\subfloat[Convergence in terms of time under IID data]{\includegraphics[width=3.0in]{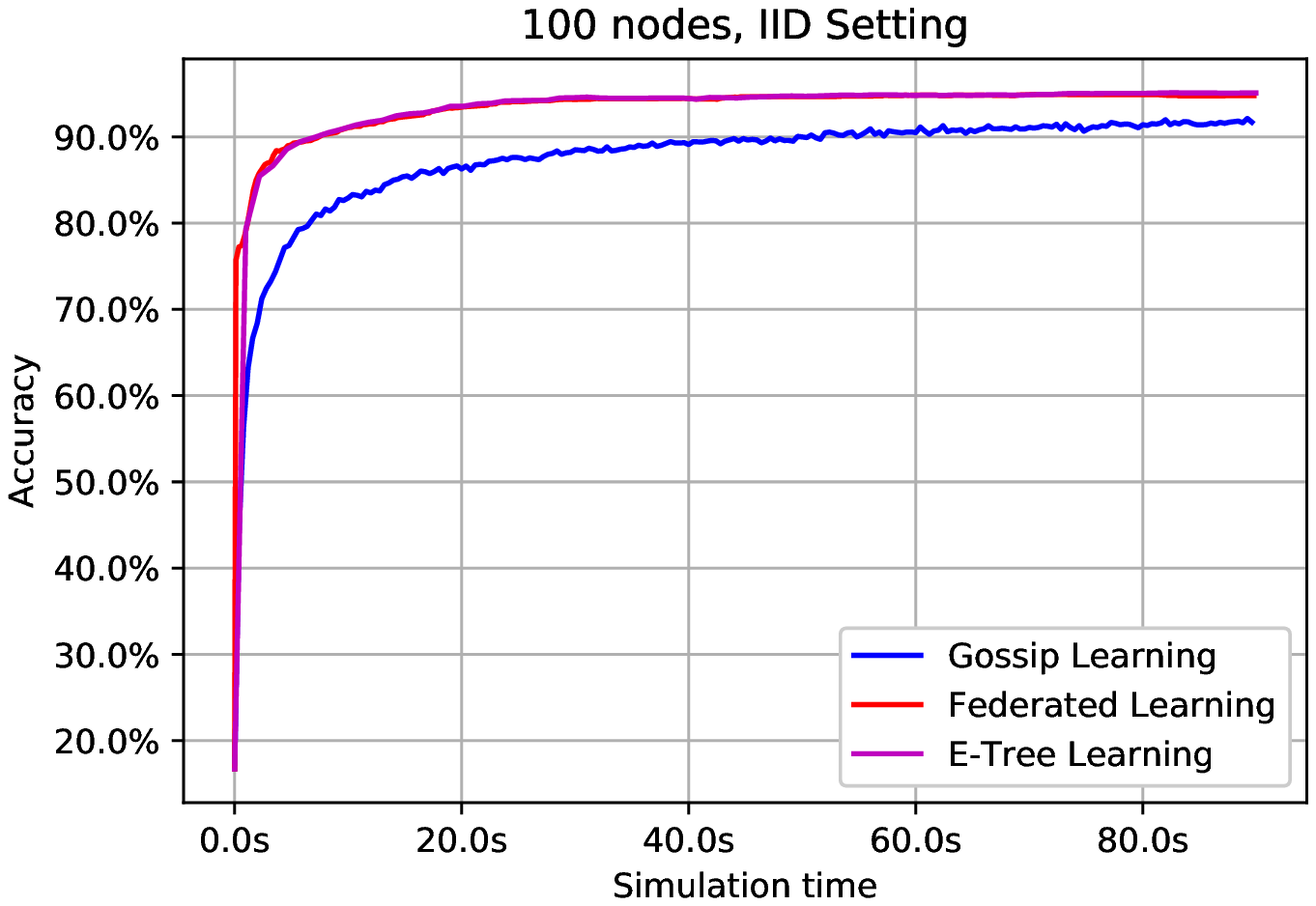}\label{conv_time_iid}}
}
\centerline{
\subfloat[Convergence in terms of round under NonIID data]{\includegraphics[width=3.0in]{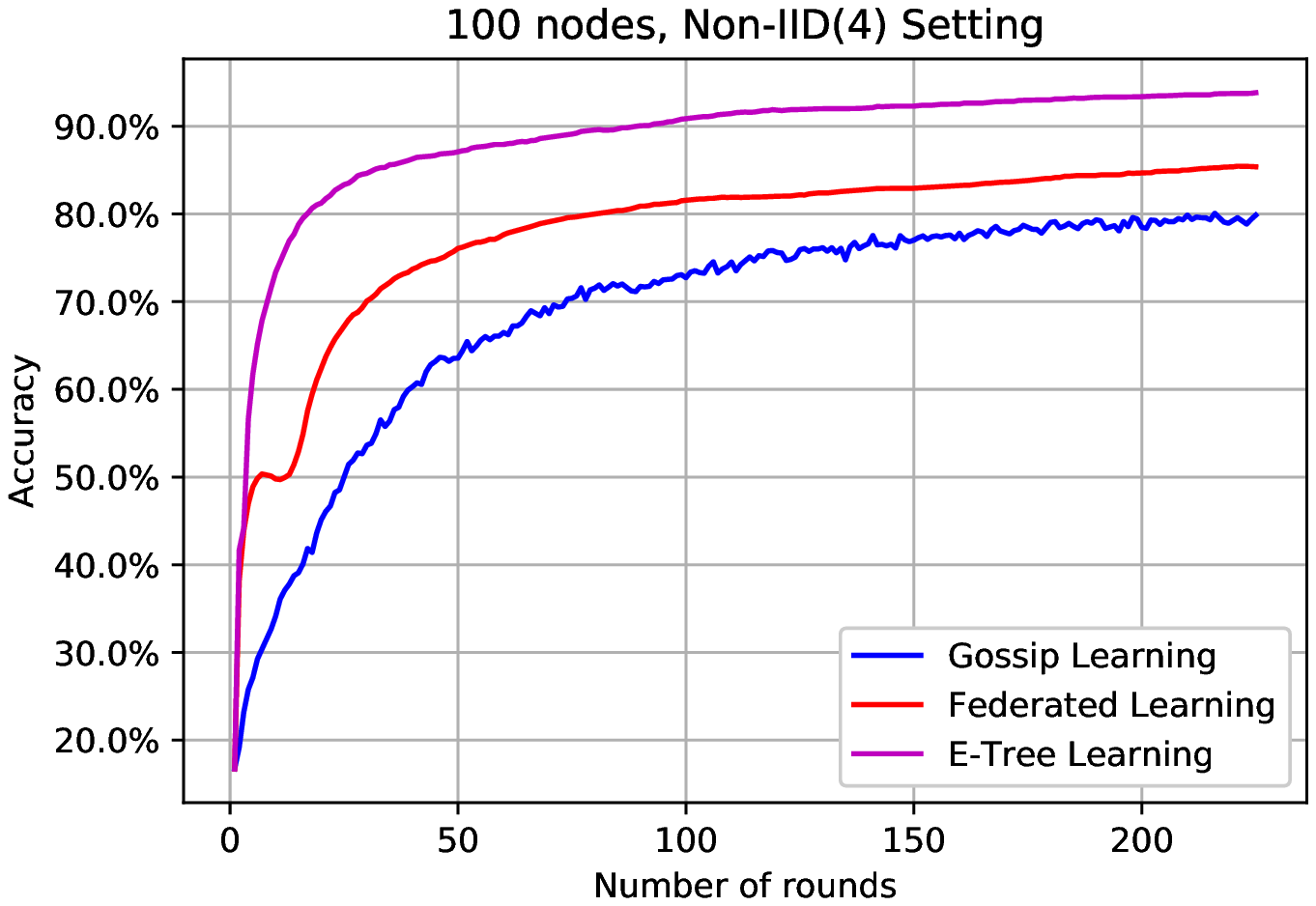}\label{conv_round}}
\hfil
\subfloat[Convergence in terms of time under NonIID data]{\includegraphics[width=3.0in]{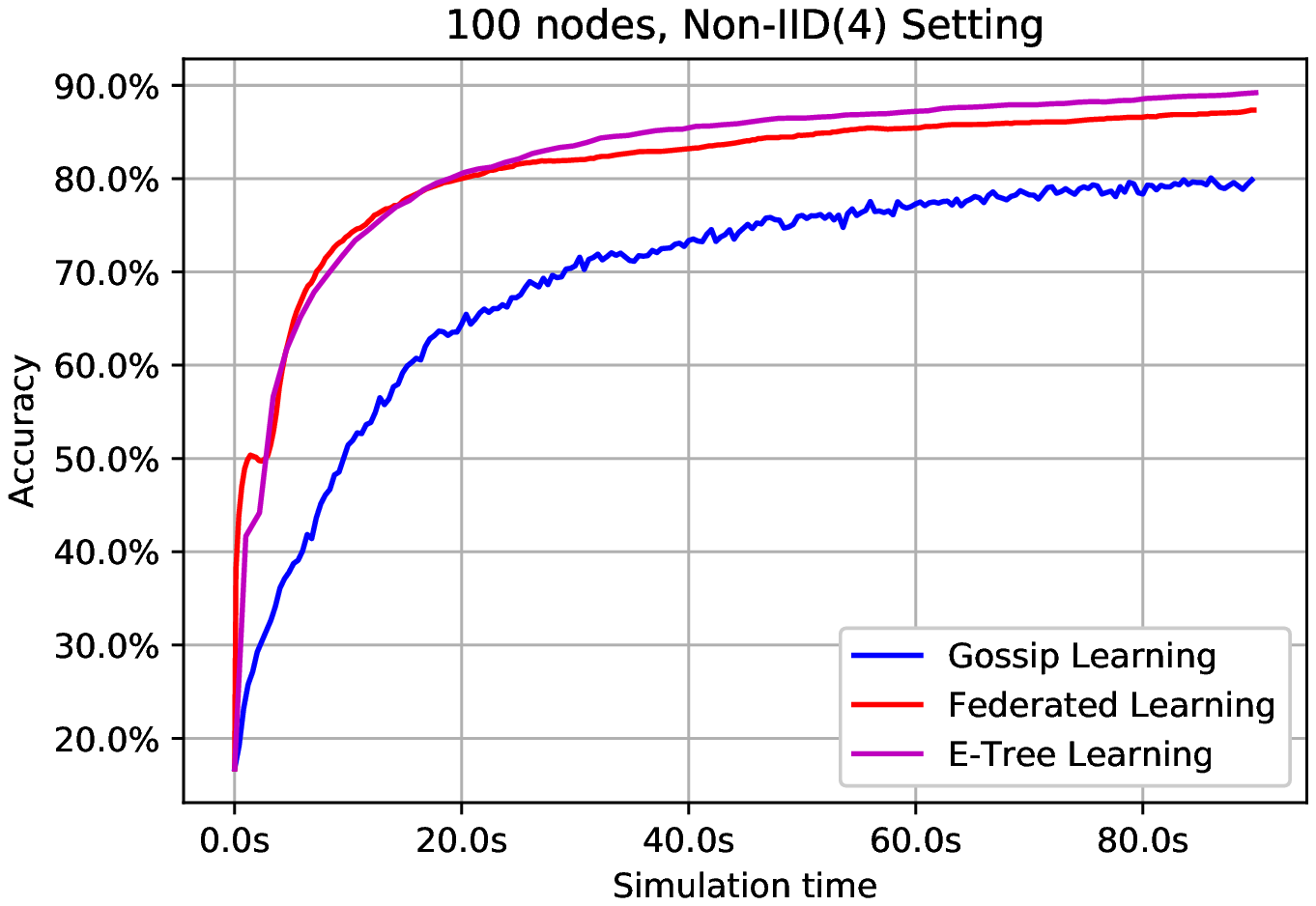}\label{conv_time}}
}
\label{fig:count}
\caption[]{Performance comparison among E-Tree learning, Federated learning and Gossip learning}\label{per_figure}
\vspace{-0.4cm}
\end{figure*}

$\bullet$ Federated learning. In the simulation, we select the centroid node in the network of edge devices as the master, while the other edge devices are the clients. The master receives model parameters from the clients, and sends them back to all the clients after model aggregation. To achieve fast convergency speed, a certain percentage $C$ of the clients would participate in every round of aggregation. By default, federated learning randomly selects the clients at each round. We set $C = 1$ since we find it is the optimal value in our simulation setting with 100 edge devices after testing various $C$. We use the shortest path routing to transmit the model parameters between the clients and master.

$\bullet$ E-Tree learning. For simplicity, we construct a three layer structure for E-Tree learning. At the bottom layer, we use k-means clustering algorithm to group the edge devices. The devices are divided into 20 groups with each group including 5 edge devices. The clustering algorithm for grouping takes into account the network distance. The centroid device in each group is selected as the aggregation node. The second layer of E-Tree has 20 aggregation nodes, where the centroid node is selected as the aggregation node at the top layer. Shortest path routing is also used to transmit model parameters from edge devices to the upper-layer aggregation node. E-Tree allows the nodes at different layers to have different aggregation frequency. In the simulation, we set the ratio of aggregation frequency between the middle layer and top layer as 5:1. It means the aggregation at the middle layer is 5 times frequent than the aggregation at the top layer.

$\bullet$ Gossip learning. Gossip learning is a fully decentralized model learning framework. In our simulation, we use the same algorithms as stated in the literature \cite{ref_8}. In the previous work, the underlying network topology is a full connective network, in which the neighbors receiving a model update are randomly selected from all the devices. In our simulation setting, we use a more practical network topology. After a local model update, the edge device sends the model to the neighbors who are physically connecting with the device.

We define two performance metrics. For federated learning and E-Tree learning, we measure the classification accuracy of the model in every round of model aggregation, where a round in E-Tree means the root node finishes an aggregation. The accuracy is calculated using the same testing dataset. Gossip learning does not have a consolidated model after a round of learning. Instead every edge device maintains a separate model. So we measure the average model accuracy of all the edge devices. We concern on the convergency of the methods. We want to compare the final and converged accuracy of the methods. This is defined as {\itshape accuracy} of the method. As well we are interested to know the {\itshape convergence time}, i.e., how much time it takes for the method to converge to a stable accuracy.

\subsection{Overall Performance Results}

Fig.\ref{conv_round_iid} and \ref{conv_round} respectively show the convergence curve of the three methods under IID and NonIID data. The model accuracy obviously increases as more rounds of aggregation are done. The three methods are able to converge to a stable accuracy. It is shown that E-Tree learning converges to a higher model accuracy than the other two methods. It also takes less number of rounds to converge to a satisfactory accuracy. We can observe that compared with case under IID data, E-Tree learning have a greater performance gain over the other methods under NonIID data. This is because E-Tree adopts a localized and grouped based aggregations considering the data properties of each device, which is able to avoid converging to a biased model.

As E-Tree learning does model aggregation at two layers, it takes more time in a round of aggregation than Federated learning. By considering the difference in round duration, Fig.\ref{conv_time_iid} and \ref{conv_time} shows the convergence speed of the methods in terms of time. From Fig.\ref{conv_time_iid}, it is shown that E-Tree has almost the same convergency speed with FL under IID. The reason is as follows. Under IID data, the advantage of E-Tree over FL is the hierarchical and localized aggregation which can significantly reduce the communication cost. We measure the communication cost as the number of hops that all the messages pass by among the devices. In the simulation, the communication cost of E-Tree is reduced by 42\% over FL. Normally in practical edge networks with high traffic with multiple applications, the reduction of communication cost would save the training time per round, because the transmission of model parameters will not cause much waiting time on the network links. However, the workloads in our simulated edge network is not high. In spite FL has high communication cost, the transmission time of model parameters among the edge device is not influenced. Thus, the convergency speed of FL is the same with E-Tree. Although the measured training time is not reduced by E-Tree, the decrease of communication cost is still meaningful, which demonstrates the advantage of E-Tree over FL.

With the same simulation setting, Fig.6d shows E-Tree can converge to a better model accuracy than FL under NonIID. Table 3 shows 2.4\% accuracy increase is obtained by E-Tree over FL. This is because that E-Tree is constructed by considering the data distribution as well as the network distance, and meanwhile the layers can have different aggregation frequency. The reason why the increase of accuracy is not great is as follows. First, the classification task and AI model may constrain the accuracy gain. From the results, we see that FL converges relatively good accuracy. The further improvement in accuracy is limited. In our future work, we will change the classification task with AI model and the data distribution in which the FL faces severe model inaccuracy. Second, as discussed in Section 6, the performance of E-Tree can be improved by further optimizing the number of layers and the aggregation frequency, which are not solved currently in this work.

\begin{table}[!t]
\renewcommand{\arraystretch}{1.1}
\centering
\caption{Accuracy of various methods}\label{result_tab}
\begin{tabular}{p{3.2cm}|l|l}
\hline\
\textbf{Methods} & \textbf{IID}  & \textbf{NonIID}  \\
\hline
Gossip Learning & 91.7\% & 79.4\% \\
\hline
Federated Learning & 94.7\% & 91.7\% \\
\hline
E-Tree Learning  & 95.0\% & 94.1\% \\
\hline
\hline
Individual Training & 81.9\% & 45.7\% \\
\hline
Grouped Aggregation & 90.1\% & 72.6\% \\
\hline
\end{tabular}
\end{table}

Table \ref{result_tab} compare the converged accuracy for the methods after sufficient training time. Under IID dataset, E-Tree has slightly higher accuracy than FL and gossip learning. However, under NonIID dataset, E-Tree has better accuracy than Federated learning and Gossip learning. This is because E-Tree adopts a grouped based localized learning mechanism. Under NonIID, the loss of a learned model highly depends on the {\itshape divergency} in data distribution of the participants. In E-Tree, the divergence would be low in a group, thus the in-group model aggregation and learning would be better than Federated learning which can be considered as only one large group of devices.

To validate the generality of the simulation setting, we measure the accuracy of two simple baseline methods. First, we want to know whether the environment setting indeed needs jointly model aggregation among the devices. We measure the model accuracy when each device individually learns the model by itself. As shown in Table \ref{result_tab}, under both data distributions, the three distributed learning frameworks improves the accuracy greatly. Especially under the NonIID data, the improvement is more significant. Second, we want to know what if E-Tree learning only contains group based aggregation at the middle layer, and the global aggregation at the top layer is deleted. We name this baseline as {\itshape Grouped Aggregation}. By comparing {\itshape Grouped Aggregation} and E-Tree, we can see that the global aggregation is necessary to improve the performance.

\begin{table}[!t]
\renewcommand{\arraystretch}{1.1}
\centering
\caption{Three experimental configurations}\label{config_tab}
\begin{tabular}{l|l|l|l|l}
\hline\
Config. & $N$  & $l$ & $K_l$ & Dataset   \\
\hline
$G_1$ & 100 & 3 & $K_1$ = 8 & HAR \\
\hline
$G_2$ & 1000 & 4 & $K_1$ = 20, $K_2$ = 5 & Pendigits \\
\hline
$G_3$ & 100 & 3 & $K_1$ = 5, 8 & HAR \\
\hline
\end{tabular}
\end{table}

\subsection{Evaluation of Device Clustering Algorithms}

We evaluate the performance of E-Tree by comparing various device clustering algorithms which are presented in Section \ref{nodeClustering}. KMA is the proposed device clustering algorithm adopted in E-Tree. We compare it with two benchmark algorithms, i.e.. K-Means and Uniform KMA. {\itshape K-Means} clusters the edge devices according to the network distance (measured by transmission delay) of the devices. It does not consider the distribution of data samples on the devices. {\itshape Uniform KMA} has an opposite idea to cluster the devices depending the data distribution. It groups the devices such that the data distribution of the groups are quite different from each other.

\begin{figure*}[htbp]
	\centering
	\subfloat[KMA with different $\delta$, $K_1 = 8$]{
		\includegraphics[scale=0.194]{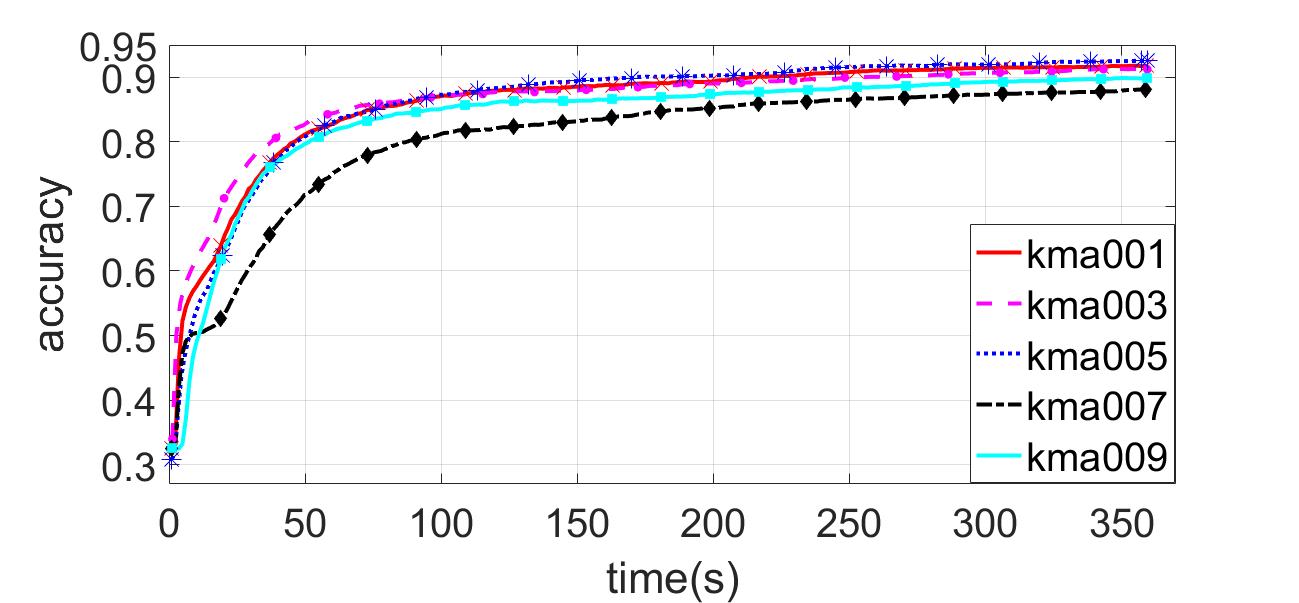}}
	\label{lc3}
	\subfloat[Different algorithm, $K_1 = 8$]{
		\includegraphics[scale=0.194]{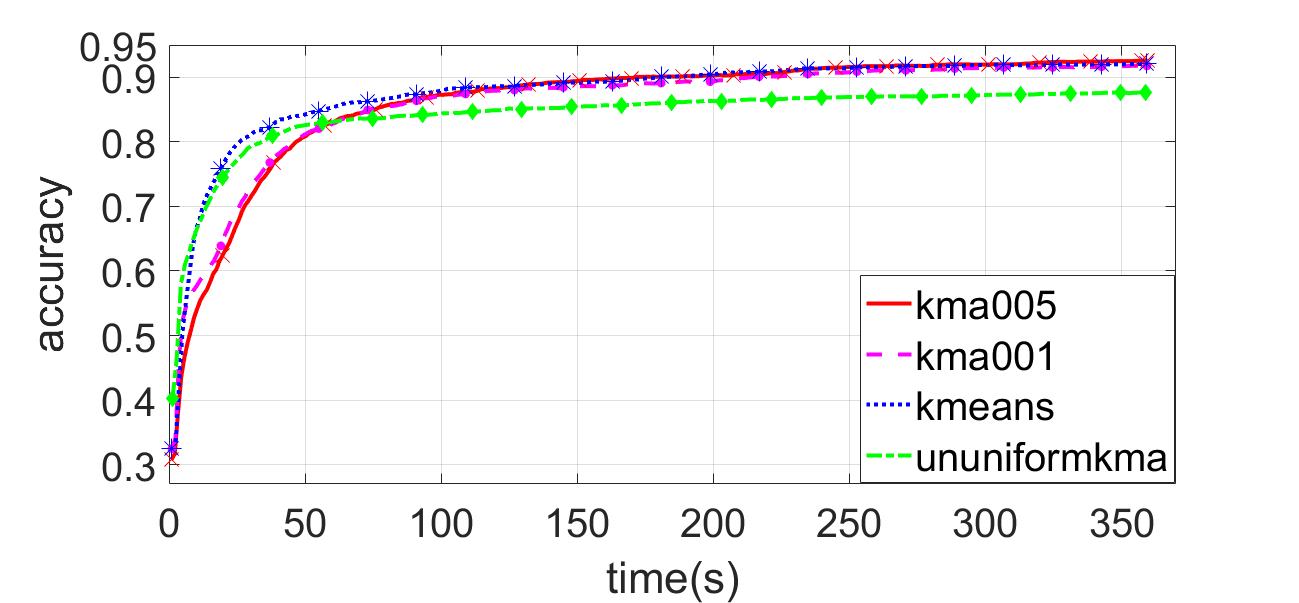}}
	\label{ld3}	
	\caption{Results of a 3-layer E-Tree with $G_1$ and HAR.}
	\label{G1_3l_HAR}
\end{figure*}

We have done the evaluation under various E-Tree configurations for different physical networks. Table \ref{config_tab} shows the configurations. For example, in the first configuration $G_1$, the physical network has 100 edge devices. We build a 3-layer E-Tree structure on the network. The number of clusters at the bottom layer is $K_1 = 8$. The dataset used in this configuration is the Human Activity Recognition (HAR). In the configuration $G_2$, the physical network scales to 1000. The E-Tree structure built on it has 4 layers, and the dataset used is Pendigits which is also from UCI machine learning repository \cite{add_11}. The data distribution is non-IID in all the experimental configurations. Specifically, we sort the data samples by class labels, and evenly assign the sorted data to nodes, such that each edge device receives data of few class labels. $D_{tes}$ in KMA contains 1000 samples selected from the original test set. The number of samples of each class is different in $D_{tes}$.

For each experiment, we first compare the accuracy of KMA algorithm by having different values of $\delta$. We observe that the proper range of $\delta$ is 0.01 to 0.1. If the values is out of this range, KMA has the same clustering result as K-Means. Hence, we compare the accuracy of KMA with a $\delta$ of 0.01, 0.03, 0.05, 0.07 and 0.09. We denote KMA with $\delta$ value of 0.01 as {\itshape KMA001}. Then, we select two optimal $\delta$ values for each experiment, and compare the accuracy of KMA with the two benchmark algorithms.

\begin{figure*}[htbp]
	\centering
	\subfloat[KMA with different $\delta$]{
		\includegraphics[scale=0.194]{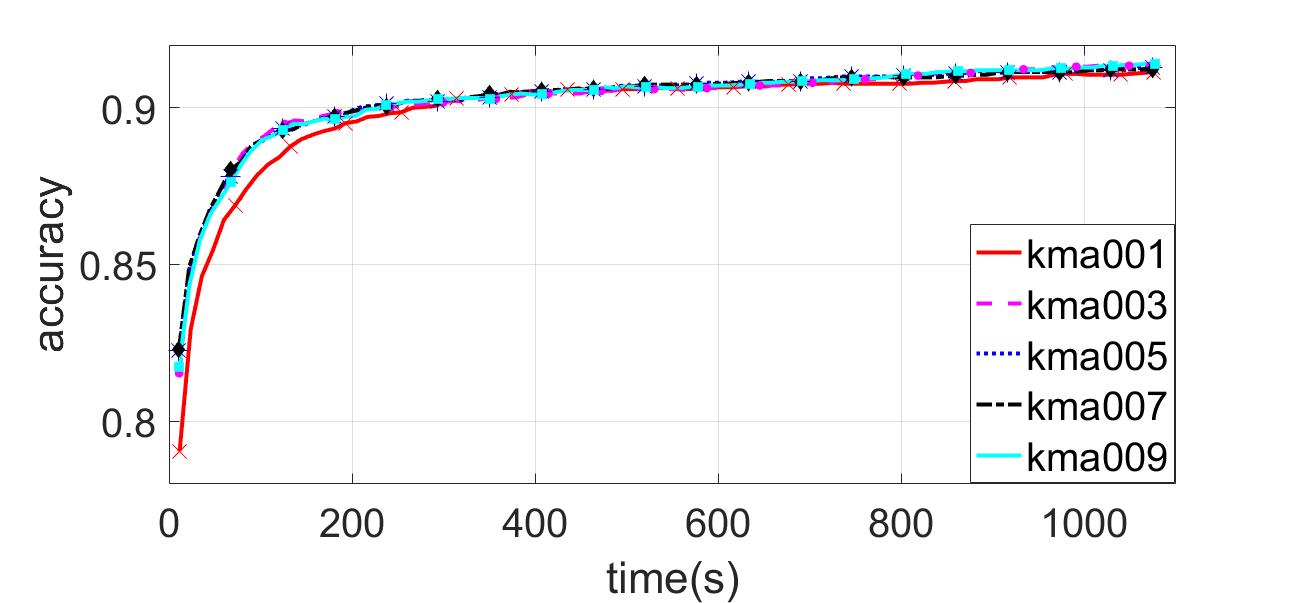}}
	\label{la5}
	\subfloat[Different algorithm]{
		\includegraphics[scale=0.194]{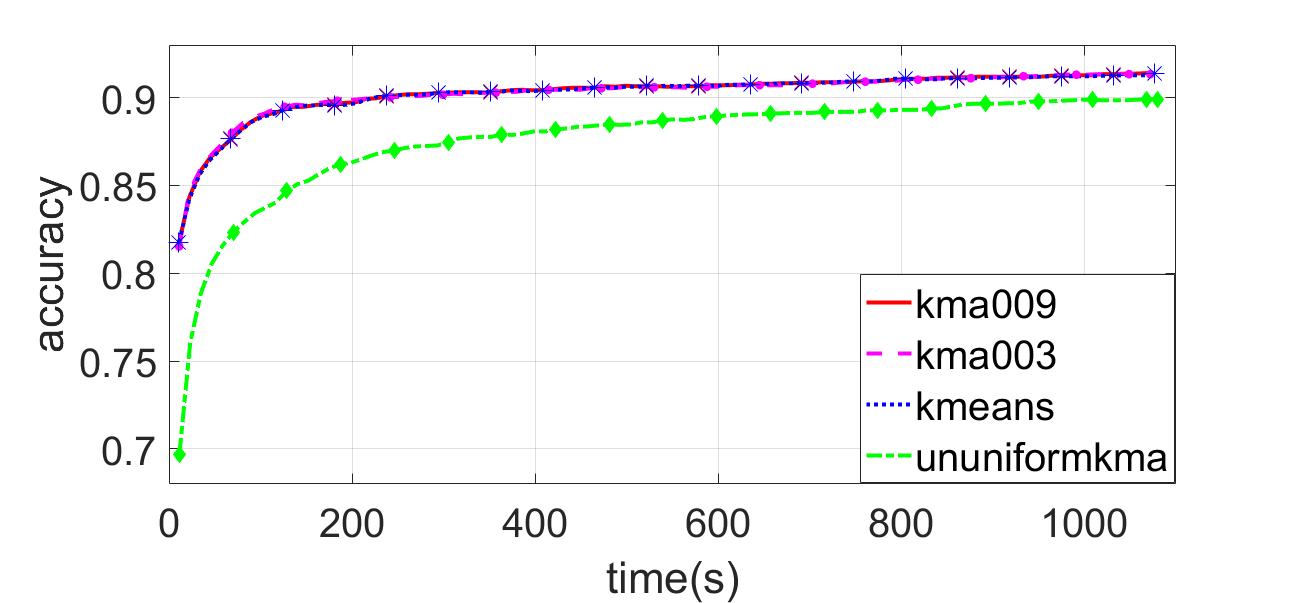}}
	\label{lb5}
	
	\caption{Results of a 4-layer E-Tree with $G_2$ and Pendigits, $K_1=20$, $K_2=5$}
	\label{G3_4l_Pendigits}
\end{figure*}

Fig.\ref{G1_3l_HAR} and \ref{G3_4l_Pendigits} shows the model accuracy of the device clustering algorithms under two network configurations. From the results, we can observe that the optimal $\delta$ varies in different conditions. Besides, KMA with optimal $\delta$ always outperforms ununiform KMA in accuracy, but the difference of accuracy between KMA and K-Means is infinitesimal in most cases. The reason is that the data samples on the devices within a small neighboring range happens to be quite diverse such that a cluster generated by grouping nearby devices together contain a relatively large number of class labels. This result is the same with the purpose of our proposed KMA.

\begin{figure*}[htbp]
	\centering
	\subfloat[KMA with different $\delta$, $K_1 = 5$]{
		\includegraphics[scale=0.194]{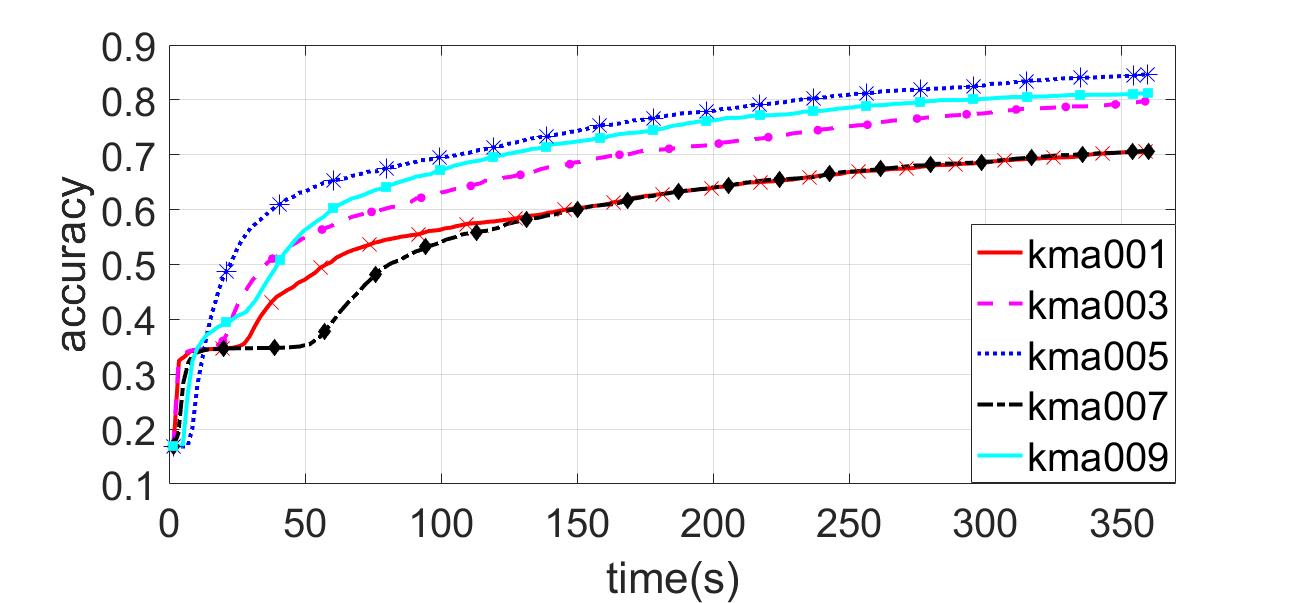}}
	\label{la6}
	\subfloat[Different algorithm, $K_1 = 5$]{
		\includegraphics[scale=0.194]{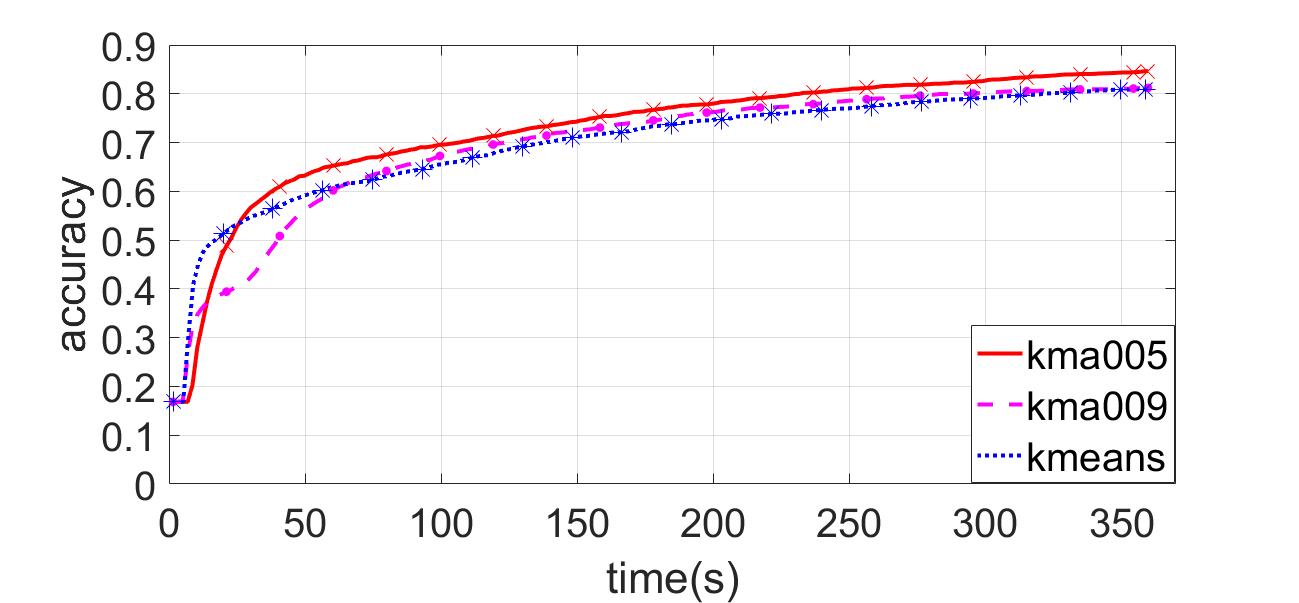}}
	\label{lb6}
	\subfloat[KMA with different $\delta$, $K_1 = 8$]{
		\includegraphics[scale=0.194]{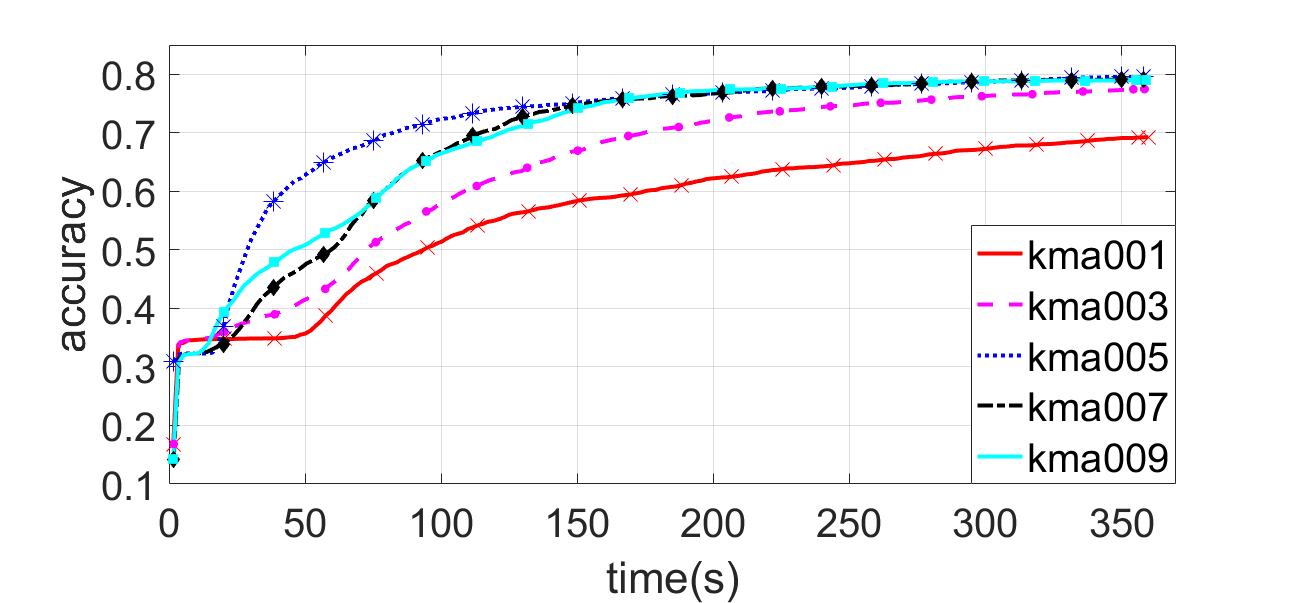}}
	\label{lc6}
	\subfloat[Different algorithm, $K_1 = 8$]{
		\includegraphics[scale=0.194]{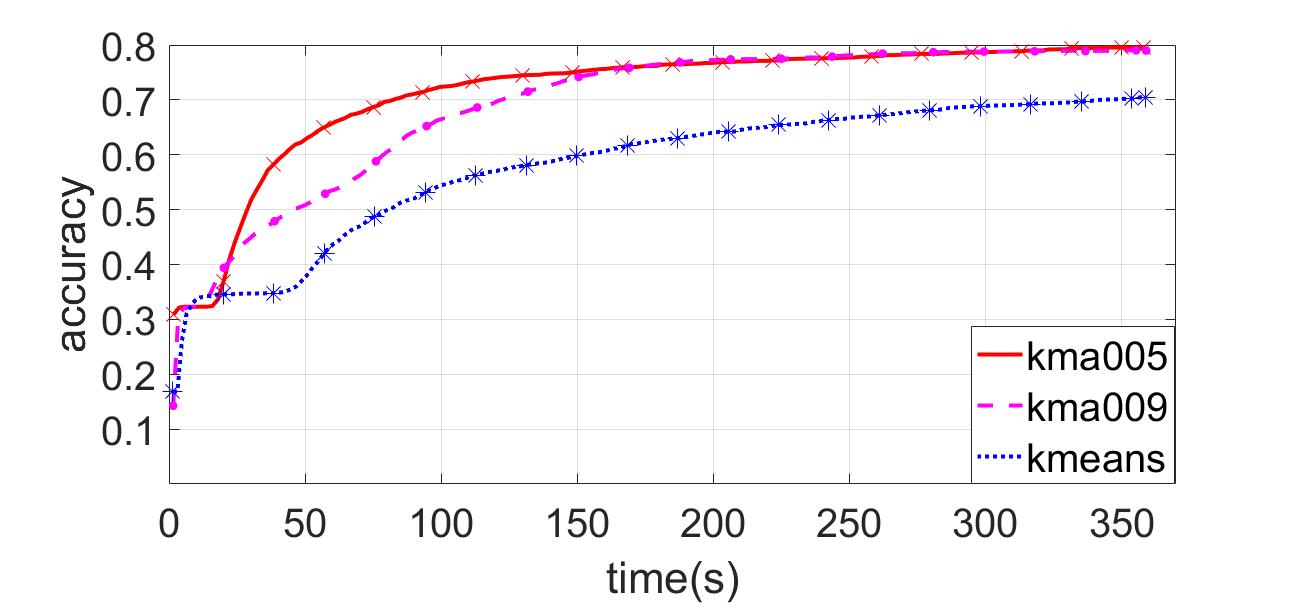}}
	\label{ld6}
	\caption{Results of a 3-layer E-Tree with $G_3$ and HAR.}
	\label{G4_3l_HAR}
\end{figure*}

We further generate an experiment configuration $G_3$ to compare the performance under more general cases where the diversity of data distribution on the devices increase with the network distance. It means the device has similar data distribution with its 1-hop neighboring devices, but has a quite different distribution from the device far away in the network. To simulate the physical network topology in $G_3$, for each class of the dataset, we randomly select a center device that owns samples of this class, and connect it to the other devices which own the same class, and set the transmission delay to a value with a mean of 50 and a variance of 10. We use HAR dataset and a 3-layer E-Tree to perform the experiments under $G_3$. The experiment results are shown in Fig.~\ref{G4_3l_HAR}. As shown in the results, although the accuracy decreases in both algorithms, KMA with optimal $\delta$ obviously outperforms K-Means in accuracy.

\section{Discussions}

\textbf{Future Work for Optimization of E-Tree}. E-Tree learning is a novel decentralized model learning framework with a tree based hierarchical structure for model aggregation. It is a more general framework than federated learning. From the evaluation results, we show that E-Tree outperforms federated learning in terms of model accuracy and convergency time in a flat network where no centralized cloud/server exists. E-Tree learning has three super-parameters that greatly affect the performance, i.e., the grouping of devices, the number of layers and the aggregation frequency of the aggregation nodes. In this paper, we have developed a new device grouping algorithm named by KMA, and compared the performance of KMA with other benchmark algorithms. In our future study, we will improve the performance of E-Tree by further optimizing the number of layers and the aggregation frequency.

\textbf{Advantages of E-Tree over Federated Learning}. Federated learning can be considered as a special structure of E-Tree. It contains two layers with the edge devices at the bottom layer and the sole aggregation node at the top layer. In a network with relatively large number of nodes, E-Tree would outperform FL by using localized aggregations to avoid high communication cost. Data aggregation has been extensively studied in Wireless Sensor Network \cite{add_16}. It is an effective method to reduce the data traffic through in-network processing and thus speed up the data collection. However, when the network scale is small, two layers maybe enough for model aggregation. In this case, E-Tree would adapt to be the same with FL.

People may argue that E-Tree is simply a federated learning with applying a tree based data aggregation, where data is the model parameters. This is not true. E-Tree does not simply aggregate the model updates from the devices to the root node using a data aggregation tree.  E-Tree allows that the aggregation frequency of each node can be different. By properly selecting the aggregation frequency, E-Tree incorporates the 'important and unique' data owned by some devices more frequently than the other data for model learning, and also fully utilizes the resources of the devices and network. Optimizing the aggregation frequency is extremely necessary under a NonIID data distribution, because in this case the edge devices contribute unequally to the model convergence. If all the nodes of E-Tree have the same aggregation frequency, E-Tree will be a federated learning simply with a data aggregation tree.

\textbf{Solving Straggler in Heterogenous Resources}. In construction of E-Tree, the device with a high processing load may become the straggler of the group and finally increase the convergency time. To solve the problem, one approach is to consider the processing load into the clustering algorithm, and try to choose the devices with similar processing load into a group. However, this approach would complicate the clustering algorithm with considering too many factors. Trade-off among these factors may influence the converged model accuracy which is dominated by the data distribution on the devices. Our approach to solve the straggler issue is that we allow the heterogeneity in terms of processing loads and computing capability within a group, while we tackle the heterogeneities by using a week synchronization approach in controlling the aggregation frequency (discussed in Section 4.2).

Week synchronization allows different aggregation frequency at various aggregation nodes in order to fully utilize the resources of the heterogeneous edge devices under various process loads. The computing capacity of the device and the processing load are important to the time of local update. The device with a short local update time does more times of local update than the other devices before a global aggregation. In this case, the straggler can be avoided on heterogeneous devices. Now the KMA algorithm for clustering the devices mainly considers the network distance (measured by transmission delay) and data distribution (measured by a pre-trained accuracy).

\section{Related Works}

Edge AI exploits edge computing infrastructures and technologies to improve the performance of AI model training and inference. There exist early surveys on this research field \cite{ref_1} \cite{ref_2}. Representative works on this field can be classified into two categories: model training and model inference. Model training uses the edge resources to efficiently train a model. Model inference focuses on enhancing the inference performance of a model by using edge computing, while the model training is still in the cloud. The closely related research to this article pertains to the model training.

Federate learning is a popular training model widely applied for Edge AI \cite{ref_16} \cite{ref_18}. Similar to the approach of parameter server in Distributed Machine Learning (DML), federate learning has a centralized model aggregation architecture where the workers/clients at the edge devices train their own models based on the local data, and a global master on the cloud aggregates the parameters from the clients. Different and more challenging then DML, federate learning has a larger number of workers involved into the training, while each client has only a small data set relative to the whole data set \cite{ref_19}. Besides, when applying federate learning into edge computing environment, more issues need to be solved such as the unbalanced and non-IID data on the clients \cite{ref_18}, resources constraints of edge devices and the networks, synchronization among the clients and so on. Zhao et al \cite{ref_5} solved the problem of non-IID data by injecting a small set of shared data to the clients for model learning. The work analyzed the trade-off between the size of the dataset injected and the performance of the learning results. Lin et al [4] proposed a method of deep gradient compression to reduce the required communication bandwidth in distributed model training. The idea is to compress the model update transmitted from the clients to the master.

To reduce the synchronization time among clients in federated learning, Nishio et al \cite{ref_3} studied the client selection problem with heterogeneous resources at the edge devices. Rather than randomly choosing the clients, this work proposed a method to select clients according to the computing capability and data size. The aggregation frequency was also optimized to reduce the synchronization time among clients in each aggregation. Wang et al \cite{ref_7} proposed a control algorithm that determines the best trade-off between local update and global parameter aggregation to minimize the loss function under a given resource budget. Lin et al \cite{add_1} studied the multi-task learning problem in the framework federated learning.

Although federated learning solves the data privacy issue by aggregating model parameters via a centralized master, it faces performance bottleneck and scalability problem when applied in Edge AI due to the resources constraints and unstable communications. Gossip learning \cite{ref_8} explored a decentralized architecture of model aggregation by removing the central aggregation master. Hardy et al \cite{ref_9} used the same gossip protocol to train a GAN model in a distributed manner. Gossip learning is a fully distributed aggregation framework with randomly updating the model parameters among the edge devices. The random aggregation does not consider the data distribution on the edge devices, so the performance in terms of the accuracy and convergency speed is not satisfactory.

In order to solve resource limitation of edge devices, some researchers investigated how to use the knowledge transfer techniques to transfer information from a large network (termed teacher) to a small one (termed student) in order to improve the performance of learning on the edge. Sharma et al \cite{add_2} studied the performance in knowledge transfer and analyzed the effectiveness by some preliminary experiments. This techniques so far is a still unexplored direction for improving the model learning performance in Edge AI.

In the perspective of model inference, related works targets on various performance metrics such as inference latency, accuracy, energy consumption, communication overhead and privacy. The main techniques used for model inference include model compression and model partition. Han et al \cite{add_3} proposed a method with weight pruning and data quantization to reduce the model complexity and resource requirement in order to enable the local inference on edge devices. The method combines difference compression techniques on demand and achieves good compression results. Kang et al \cite{add_4} designed an architecture that supports a partitioned execution of model inference between the edge and cloud. This work selected the optimal partition point at a multi-layer neural network via up front performance prediction of each layer. Li et al \cite{add_5} set many exist points and allow the model inference terminated earlier at the exist point in order to reduce the latency. They proposed a method to select the best partition point and exit point of a DNN model. Drolia et al \cite{add_6} proposed to cache and reuse the inference results with the aim of reducing the inference latency. In order to balance the accuracy, latency and energy, Taylor et al \cite{add_7} trained a set of models for a task with various model size, and selected the model adaptively for inference.

\section{Conclusion}

In this article we explore the distributed machine learning frameworks for Edge AI. We propose E-Tree learning, which is a novel learning framework with decentralized model aggregations in edge computing networks. E-Tree leverages a tree structure to perform a localized and level by level model aggregation from the edge devices. We present the general model and individual changing issues including the device clustering, aggregation frequency controlling, scheduling and so on. We compare the performance of E-Tree with federated learning and Gossip learning using an Open Source benchmark simulator. Although some superparameters of E-Tree are not tuned to the optimum, E-Tree still outperforms the federated learning by 2.4\% in accuracy and Gossip learning by 14.7\% under a NonIID data distribution. Results shows that E-Tree has faster convergency speed than the two benchmark frameworks as well. We also evaluated the proposed KMA device clustering algorithm. The results show that the KMA algorithm can significantly improve the model accuracy of E-Tree by clustering the devices based data distribution as well as the network distanace.

% or


\ifCLASSOPTIONcompsoc
  \section*{Acknowledgments}
\else
  \section*{Acknowledgment}
\fi

This work is supported in part by the National Natural Science Foundation of China under Grant No.61972161, in part by Hong Kong RGC General Research Fund under Grant PolyU 152133/18 and PolyU 15217919, in part by Guangdong Basic and Applied Basic Research Foundation under Grant 2020A1515011496.

\ifCLASSOPTIONcaptionsoff
  \newpage
\fi



%\bibliographystyle{IEEEtran}

\begin{thebibliography}{1}

\bibitem{ref_7}
{S. Wang, T. Tuor and et al}.
\newblock {Adaptive Federated Learning in Resource Constrained Edge Computing Systems}.
\newblock In {\em {IEEE Journal on Selected Areas in Communications}}, vol.37, no.6, pp.1205-1221, June 2019.

\bibitem{ref_36}
{M. Li, D. Andersen and et al}.
\newblock {Communication Efficient Distributed Machine Learning with the Parameter Server}.
\newblock In {\em {Proc. of NeurIPS}}, pp.19-27, 2014.

\bibitem{ref_8}
{I. Hegedus, G. Danner and et al}.
\newblock {Gossip Learning as a Decentralized Alternative to Federated Learning}.
\newblock In {\em {In Proc. DAIS 2019 and LNCS}}, pp.74–90, 2019.

\bibitem{ref_1}
{Z. Zhou, X. Chen and et al}.
\newblock {Edge Intelligence: Paving the Last Mile of Artiﬁcial Intelligence With Edge Computing}.
\newblock In {\em {Proceedings of the IEEE}}, vol.107, no.8, pp.1738-1762, 2019.

\bibitem{ref_2}
{T. Rausch and S. Dustdar}.
\newblock {Edge Intelligence: The Convergence of Humans, Things, and AI}.
\newblock In {\em {Proc. of IEEE International Conference on Cloud Engineering}}, pp.86-96, 2019.

\bibitem{ref_16}
{N. Tran, W. Bao and et al}.
\newblock {Federated Learning over Wireless Networks: Optimization Model Design and Analysis}.
\newblock In {\em {Proc. of INFOCOM}}, June 2019.

\bibitem{ref_18}
{X. Li, K. Huang and et al}.
\newblock {On the Convergence of FedAvg on Non-IID Data}.
\newblock In {\em {arXiv:1907.02189v2}}, October 2019.

\bibitem{ref_19}
{K. Bonawitz, H. Eichner and et al}.
\newblock {Towards Federated Learning at Scale: System Design}.
\newblock In {\em {arXiv:1902.01046v2}}, March 2019.

\bibitem{ref_5}
{Y. Zhao, M. Li and et al}.
\newblock {Federated Learning with Non-IID Data}.
\newblock In {\em {arXiv:1806.00582}}, June 2018.


\bibitem{ref_4}
{Y. Lin, S. Han and et al}.
\newblock {Deep Gradient Compression: Reducing the Communication Bandwidth for Distributed Training}.
\newblock In {\em {Proc. of ICLR 2018}}.

\bibitem{ref_3}
{T. Nishio, R. Yonetani and et al}.
\newblock {Client Selection for Federated Learning with Heterogeneous Resources in Mobile Edge}.
\newblock In {\em {Proc. of ICC}}, May 2019.

\bibitem{ref_7}
{S. Wang, T. Tuor and et al}.
\newblock {Adaptive Federated Learning in Resource Constrained Edge Computing Systems}.
\newblock In {\em {IEEE Journal on Selected Areas in Communications}}, vol.37, no.6, pp.1205-1221, June 2019.

\bibitem{ref_9}
{C. Hardy, E. Merrer and et al}.
\newblock {Gossiping GANs}.
\newblock In {\em {Proc. of DIDL}}, pp.25-28, December 2018.

\bibitem{add_1}
{V. Smith, C. Chiang and et al}.
\newblock {Federated Multi-Task Learning}.
\newblock In {\em {Proc. of NIPS}} 2017.

\bibitem{add_2}
{R. Sharma, S. Biookaghazadeh and et al}.
\newblock {Are Existing Knowledge Transfer Techniques Effective for Deep Learning with Edge Devices?}
\newblock In {\em {Proc. of IEEE International Conference on Edge Computing}} 2018.

\bibitem{add_3}
{S. Han, H. Map and et al}.
\newblock {Deep compression: Compressing Deep Neural Networks with Pruning, Trained Quantization and Huffman Coding}.
\newblock In {\em {Proc. of ICLR}} 2016.

\bibitem{add_4}
{Y. Kang, J. Hauswald and et al}.
\newblock {Neurosurgeon: Collaborative Intelligence Between the Cloud and Mobile Edge}.
\newblock In {\em {Proc. of ASPLOS}} 2017.

\bibitem{add_5}
{E. Li, Z. Zhou and et al}.
\newblock {Edge Intelligence On-Demand Deep Learning Model Co-Inference with Device-Edge Synergy}.
\newblock In {\em {arXiv}} 2018.

\bibitem{add_6}
{U. Drolia, K. Guo and et al}.
\newblock {Cachier: Edge-caching for Recognition Applications}.
\newblock In {\em {Proc. of ICDCS}} 2017.

\bibitem{add_7}
{B. Taylor, S. Marco and et al}.
\newblock {Adaptive Deep Learning Model Selection on Embedded Systems}.
\newblock In {\em {ACM SIGPLAN Notices}}, vol.53, no.6, pp.31-43.

\bibitem{add_8}
{Y. Jin, J. Jin and et al}.
\newblock {An Intelligent Task Allocation Scheme for Multiple Wireless Networks}.
\newblock In {\em {IEEE Transactions on Parallel and Distributed Systems}}, vol.23, no.3, pp.444-451, 2012


\bibitem{add_9}
{H. Brendan McMahan, E. Moore and et al}.
\newblock {Communication-efficient Learning of Deep Networks from Decentralized Data}.
\newblock In {\em {arXiv:1602.05629v3}}, Feb., 2017


\bibitem{add_10}
{M. Abad, E. Ozfatura and et al}.
\newblock {Hierarchical Federated Learning across Heterogeneous Cellular Networks}.
\newblock In {\em {2020 IEEE International Conference on Acoustics, Speech and Signal Processing }}, May, 2020


\bibitem{add_11}
{D. Dua, C. Graff and et al}.
\newblock {UCI Machine Learning Repository}.
\newblock In {\em {http://archive.ics.uci.edu/ml}}, 2019

\bibitem{add_14}
{Y. Xiong, Y. Sun and et al}.
\newblock {Extend Cloud to Edge with KubeEdge}.
\newblock In {\em {Proc. of ACM SEC}}, Octber 2018

\bibitem{add_12}
{N. Abbas, Y. Zhang and et al}.
\newblock {Mobile Edge Computing: A Survey}.
\newblock In {\em {IEEE Internet of Things Journal}}, vol.5, no.1, 2018

\bibitem{add_13}
{Y. Sahni, J. Cao and et al}.
\newblock {Mobile Edge Computing: A Survey}.
\newblock In {\em {Edge Mesh: A New Paradigm to Enable Distributed Intelligence in Internet of Things}}, vol.5, 2017

\bibitem{add_15}
{S. Deng, H. Zhao and et al}.
\newblock {Edge Intelligence: The Confluence of Edge Computing and Artificial Intelligence}.
\newblock In {\em {IEEE Internet of Things Journal}}, vol.7, no.8, 2020

\bibitem{add_16}
{R. Rajagopalan and P. Varshney}.
\newblock {Data-aggregation Techniques in Sensor Networks: A survey.}.
\newblock In {\em {IEEE Communications Surveys \& Tutorials}}, vol.8, no.4, 2006

\bibitem{add_17}
{R. Rajagopalan and P. Varshney}.
\newblock {Minimizing Resource Consumption Cost of DAG Applications With Reliability Requirement on Heterogeneous Processor Systems}.
\newblock In {\em {IEEE Transactions on Industrial Informatics}}, vol.16, no.12, 2020

\bibitem{add_18}
{R. Rajagopalan and P. Varshney}.
\newblock {Reliability-driven Scheduling of Parallel Real-Time Jobs in Heterogeneous Systems}.
\newblock In {\em {Proc. of ICPP}}, September, 2001

\end{thebibliography}

\begin{IEEEbiography}[{\includegraphics[width=1in,height=1.25in,clip,keepaspectratio]{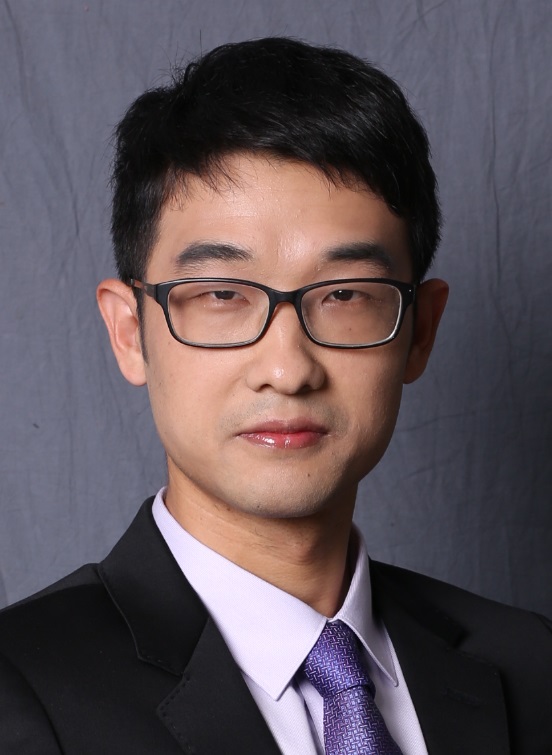}}]%
{Lei Yang} is currently an associate professor at the School of Software Engineering, South China University of Technology, China. He received the BSc degree from Wuhan University, in 2007, the MSc degree from the Institute of Computing Technology, Chinese Academy of Sciences, in 2010, and the PhD degree from the Department of Computing, Hong Kong Polytechnic University, in 2014. He has been a visiting scholar at Technique University Darmstadt, Germany from Nov. 2012 to Mar. 2013. His research interests include edge and cloud computing, distributed machine learning, and scheduling and optimization theories and techniques.
\end{IEEEbiography}
\vspace{-0.3cm}

\begin{IEEEbiography}[{\includegraphics[width=1in,height=1.25in,clip,keepaspectratio]{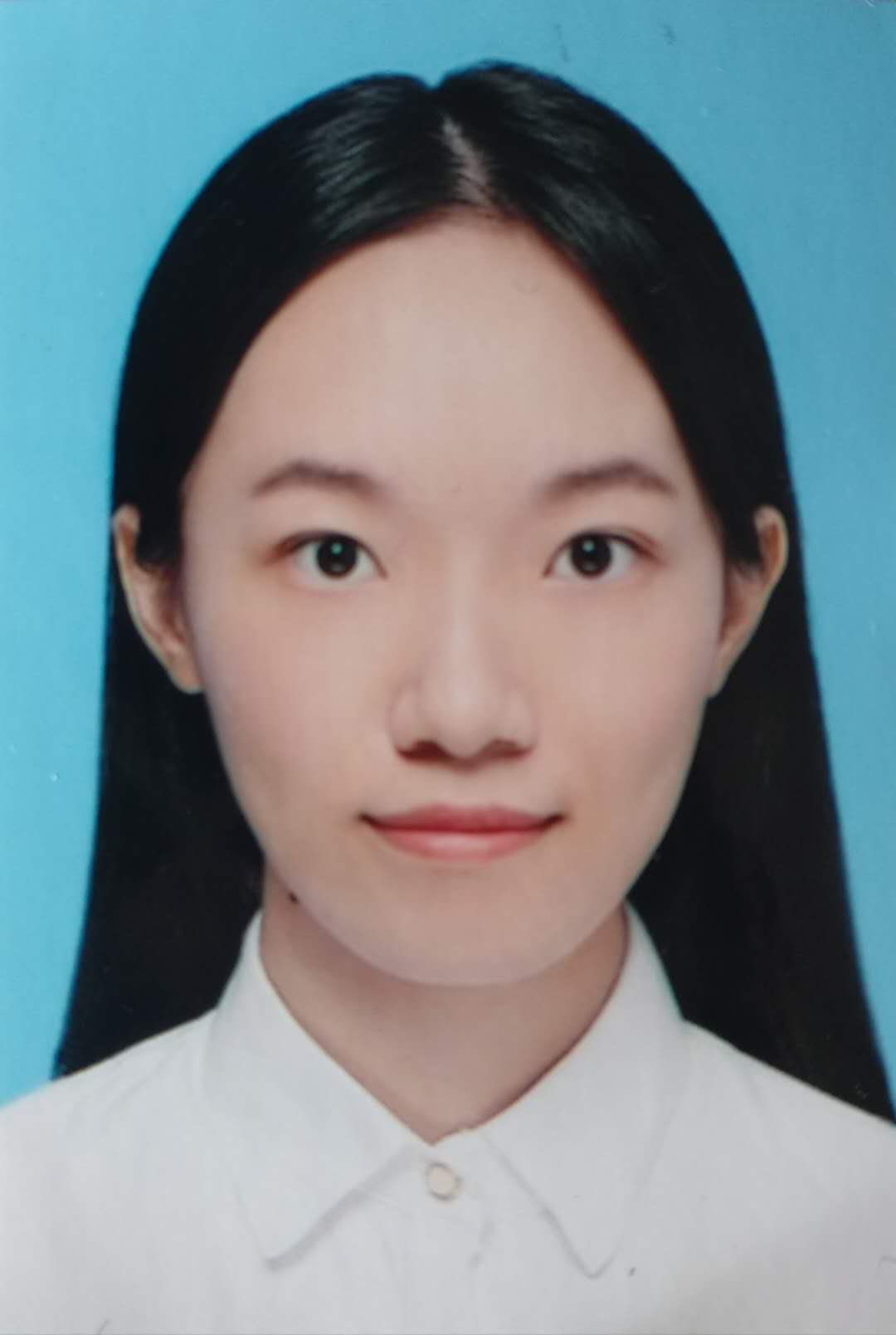}}]%
{Yanyan Lu} is a 1st year postgraduate student at School of Software Engineering, South China University of Technology, China, where she received the B.Eng. degree in software engineering in June 2020. Her research interests lie in distributed machine learning, edge computing and intelligence.
\end{IEEEbiography}
\vspace{-0.3cm}

\begin{IEEEbiography}[{\includegraphics[width=1in,height=1.25in,clip,keepaspectratio]{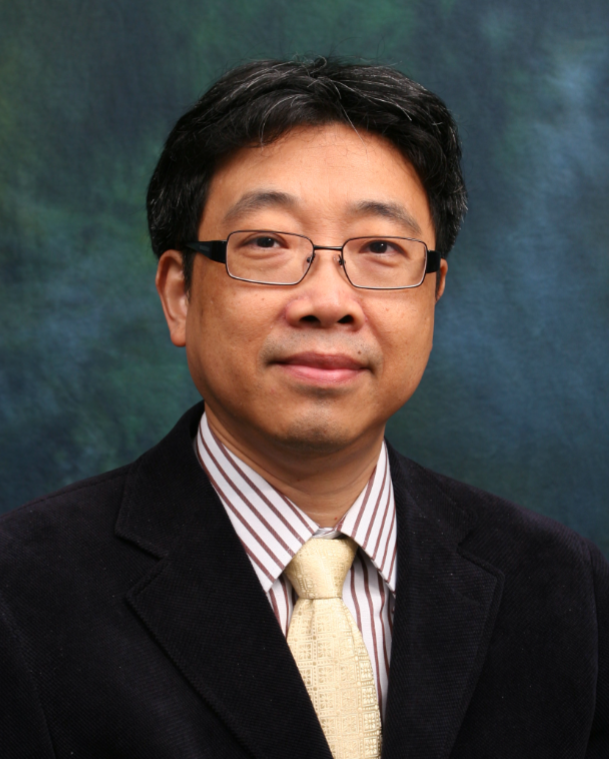}}]%
{Jiannong Cao} is a Chair Professor of Distributed and Mobile Computing of the Department of Computing at The Hong Kong Polytechnic University. He is also the director of the Internet and Mobile Computing Lab in the department and the director of University Research Facility in Big Data Analytics. He received the B.Sc. degree in computer science from Nanjing University, China, in 1982, and the M.Sc. and Ph.D. degrees in computer science from Washington State University, USA, in 1986 and 1990 respectively. His research interests include parallel and distributed computing, wireless networks and mobile computing, big data and cloud computing, pervasive computing, and fault tolerant computing. He has co-authored 5 books in Mobile Computing and Wireless Sensor Networks, co-edited 9 books, and published over 500 papers in major international journals and conference proceedings. He is a fellow of IEEE.
\end{IEEEbiography}
\vspace{-0.3cm}

\begin{IEEEbiography}[{\includegraphics[width=1in,height=1.25in,clip,keepaspectratio]{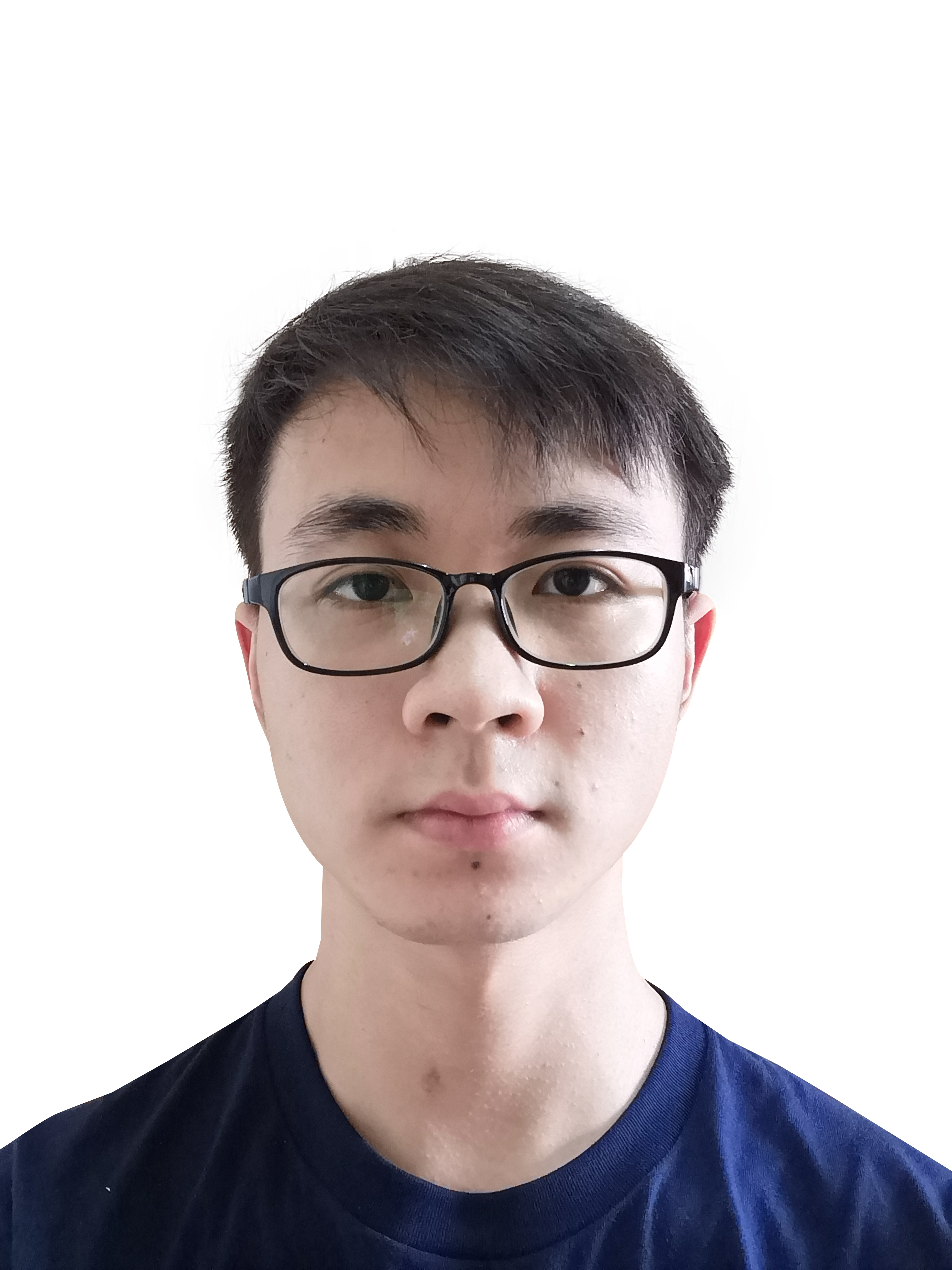}}]%%
{Jiaming Huang} is a 2nd year postgraduate student at School of Software Engineering, South China University of Technology, China. He obtained the B.Eng. degree in computer sciences from Nanchang University, China, in June 2019. His research interests are edge computing and distributed machine learning.
\end{IEEEbiography}
\vspace{-0.3cm}

\begin{IEEEbiography}[{\includegraphics[width=1in,height=1.25in,clip,keepaspectratio]{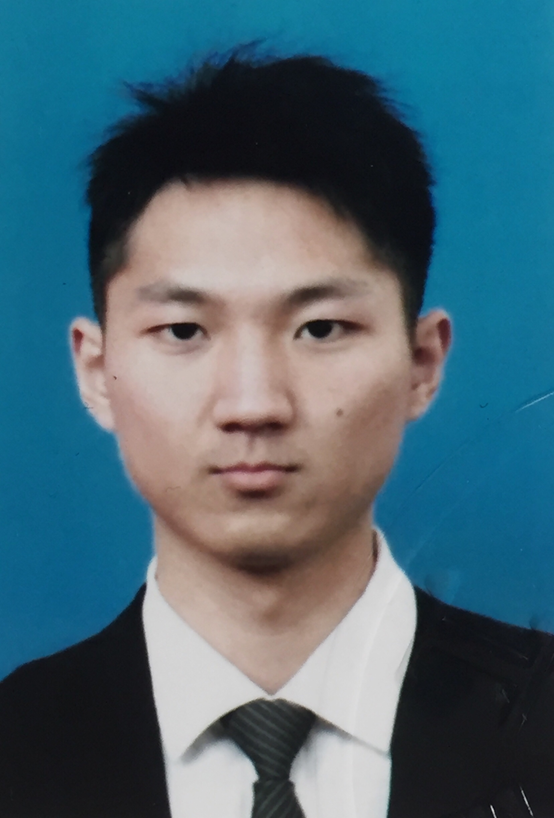}}]%
{Mingjin Zhang} is currently a Ph.D. student with the Department of Computing, The Hong Kong Polytechnic University, Hong Kong. He received the B.Eng. degree in communication engineering from Wuhan University of Technology, China, in 2019. His research interests include edge computing, distributed machine learning and Internet of Things.
\end{IEEEbiography}
\vspace{-0.3cm}


%\vfill


\end{document}